\documentclass[journal]{IEEEtran}
\usepackage{amsmath}
\usepackage{amssymb}
\usepackage{enumerate}
\usepackage{commath}
\usepackage{color}
\usepackage{graphicx}
\usepackage{multirow}
\usepackage{bigstrut}
\usepackage{tabularx}
\usepackage[flushleft]{threeparttable}
\usepackage{mathtools}
\usepackage{multicol}

\usepackage{subfigure}

\usepackage{booktabs}

\DeclareMathOperator*{\argmax}{argmax}

\ifCLASSINFOpdf
\else
\fi
\begin{document}

\bstctlcite{IEEEexample:BSTcontrol}

\title{Analysis of Rolling Shutter Effect on ENF based Video Forensics}

\author{Saffet~Vatansever,~
        Ahmet~Emir~Dirik,~
        Nasir Memon% <-this % stops a space
\thanks{This material is based on research sponsored by DARPA and the Air Force Research Laboratory (AFRL) under agreement number FA8750-16-2-0173. The U.S. Government is authorized to reproduce and distribute reprints for Governmental purposes notwithstanding any copyright notation thereon. The views and conclusions contained herein are those of the authors and should not be interpreted as necessarily representing the official policies or endorsements, either expressed or implied, of DARPA and the Air Force Research Laboratory (AFRL) or the U.S. Government.}        
\thanks{S. Vatansever is with the Department
of Mechatronics Engineering, Bursa Technical University , Bursa 16333, Turkey,
and also with Uludag University, Bursa 16059, Turkey.}
\thanks{A. E. Dirik (Corresponding Author) is with the Department of Computer Engineering, Faculty of Engineering, Uludag University, Bursa 16059, Turkey (e-mail: edirik@uludag.edu.tr).}
\thanks{N. Memon is with the Department of Computer Science and Engineering, Tandon School of Engineering, New York University, Brooklyn, NY 11201 USA.}}

\maketitle

\begin{abstract}
ENF is a time-varying signal of the frequency of mains electricity in  a power grid. It continuously fluctuates around a nominal value (50/60 Hz) due to changes in supply and demand of power over time. Depending on these ENF variations, the luminous intensity of a mains-powered light source also fluctuates. These fluctuations in luminance can be captured by video recordings. Accordingly, ENF can be estimated from such videos by analysis of steady content in the video scene. When videos are captured by using a rolling shutter sampling mechanism, as is done mostly with CMOS cameras, there is an idle period between successive frames. Consequently,  a number of illumination samples of the scene are effectively lost due to the idle period. These missing samples affect ENF estimation, in the sense of the \textcolor{black}{frequency} shift caused and the power attenuation that results. This work develops an analytical model for videos captured using a rolling shutter mechanism. The model illustrates how the \textcolor{black}{frequency of the main ENF harmonic} varies depending on the idle period length, and how the power of \textcolor{black}{the captured} ENF attenuates as idle period increases. Based on this, \textcolor{black}{a novel idle period estimation method for potential use in camera forensics} that is able to operate independently of video frame rate is proposed. Finally, a novel time-of-recording verification approach based on use of multiple \textcolor{black}{ENF components}, idle period assumptions and interpolation of missing ENF samples is also proposed.

%In literature, mainly two different ENF estimation approach is introduced for videos, which are based on global shutter and rolling shutter

\end{abstract}


% \fi
\IEEEpeerreviewmaketitle

\section{Introduction}

\IEEEPARstart{E}{NF} (Electric Network Frequency) oscillates instantaneously between an upper and lower bound around a nominal value (50/60 Hz) owing to a continuous imbalance between generated power and consumed power \cite{Bollen2006}. In an interconnected mains power network, the temporal variation of ENF  is expected to be the same across the whole region that  the network spans \cite{Grigoras2005}.

ENF is inherently embedded in audio recordings due to the electromagnetic field or the acoustic mains hum in the \textcolor{black}{environment,} and can be extracted from audio with time-domain or frequency-domain techniques \cite{Grigoras2005}, \cite{Grigoras2009}, \cite{Chai2013} \cite{Fechner2014}. In recent years, it was discovered that ENF is also embedded in video recordings of a scene illuminated by a  mains-powered light source \cite{Garg2011}. The illumination intensity of a mains-powered light source varies depending on ENF fluctuations in the power network and these variations in luminance can be measured from video recordings. Specifically, ENF can be extracted from  videos by doing a careful analysis of subtle illumination alterations in steady content through consecutive video frames \cite{Garg2011}, \cite{MinWu-SeeingENFPower-signature-basedtimestampfordigitalmultimediaviaopticalsensingandsignalprocessing}, \cite{su2014}, \cite{patent:20150356992}, \cite{Vatansever2017_SPL}.

Since the ENF signal measured from any point in the power network can be used as a reference signal, as well as the fact that it can be extracted from digital media files has led the use of ENF in media forensics. Specifically, it can be exploited for a number of forensic and anti-forensic applications including time-of-recording verification \cite{Fechner2014, MinWu-SeeingENFPower-signature-basedtimestampfordigitalmultimediaviaopticalsensingandsignalprocessing, Bykhovsky2013}, media authentication \cite{Hua2016, Savari2016}, multimedia synchronisation \cite{MinWu-ENFSignalInducedbyPowerGridANewModalityforVideoSynchronization}, \cite{MinWu-ExploringtheuseofENFformultimediasynchronization}, \cite{Chuang2013}, power grid identification \cite{MinWu-ENF-BasedRegion-of-RecordingIdentificationforMediaSignals} and \textcolor{black}{camera read-out time estimation} \cite{Hajj-Ahmad2016-ENF}.

%\textcolor{blue}{As light source flickers at both the positive and negative cycles of AC current, the illumination frequency becomes double the mains power frequency. Accordingly, the illumination signal can be treated as the absolute form of the cosine function in $(\ref{Eq_PowerGridVoltage})$. For example in Europe, nominal ENF in any region is 50 Hz, thus frequency of illumination varies around 100 Hz. According to the Nyquist Sampling Theorem, a sampling rate of at least 200 Hz is needed in order to extract illumination frequency accurately from sampled data. Although most consumer cameras are unable to provide such high frame sampling rates, it is still possible to estimate illumination frequency from its alias frequency.}

Most consumer cameras equipped with CMOS sensors use a rolling shutter sampling mechanism to capture a video frame whereas those with CCD sensors employ a global shutter sampling mechanism. When using a  global shutter mechanism, each pixel, so each row of a frame is exposed at the same time instance. Whereas with a rolling shutter, each row is captured at different time instances. This difference in the sampling mechanisms has led researchers to develop different approaches for ENF signal estimation from videos. The first approach \cite{Garg2011} is aimed at the global shutter mechanism. It is based on averaging all the steady pixels in each frame of the video, i.e.\textcolor{black}{,} one illuminations sample per frame. Consequently, it is unable to satisfy Nyquist criteria \cite{Proakis-DigitalSignalProcessingPrinciplesAlgorithmsAndApplications} and so is dependent on the alias ENF caused by the low sampling frequency as determined by the \textit{video frame rate}. Another ENF estimation approach \cite{Vatansever2017_SPL} is based on averaging steady super-pixels rather than all steady pixels within the frame, which leads the technique to be applicable to videos exposed in different sampling mechanisms. Nevertheless, since one illumination sample only is obtained per frame, this technique also relies on alias ENF. The other approach is designed for the rolling shutter mechanism and is based on averaging steady pixels in each row, i.e.\textcolor{black}{,} one illumination sample per row \cite{su2014}, \cite{patent:20150356992}. This method uses the advantage of the increased ENF sampling frequency of the rolling shutter mechanism, as determined by the \textit{video frame rate $\times$ number of rows}. However, the rolling shutter approach brings with it the issue of idle time between two successive frames where there is no sampling done. Accordingly, some illumination samples are lost at the end of each frame. Su et al. provide a fundamental understanding on how the \textcolor{black}{main ENF harmonic} is shifted to some other frequencies due to the idle period in videos captured using the rolling shutter mechanism \cite{su2014}. They use a filter bank model formulation  of the mechanism for their analysis.

%\begin{figure}[t]
%  \centering
%  \subfigure[]{\includegraphics[width=75mm]{GlobalShutter_v4}}
%%  \hfill
%  \quad
%  \subfigure[]{\includegraphics[width=75mm]{RollingShutter_v3}}
%  \caption{Demonstration of (a) Global Shutter Sampling Mechanism (b) Rolling Shutter Sampling Mechanism for a single frame}
%  \label{Fig: Global Shutter vs Rolling Shutter}
%\end{figure}

\begin{figure}[t]
  \centering
  \includegraphics[width=75mm]{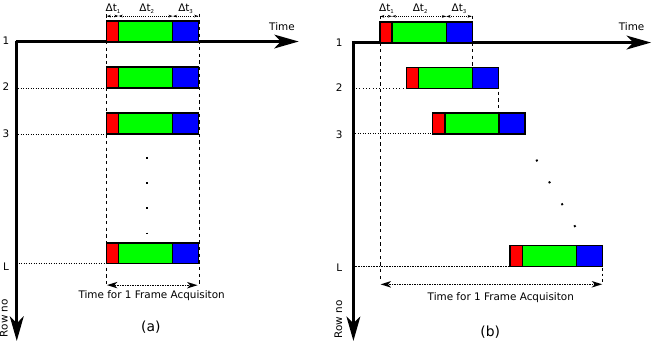}
  \caption{Demonstration of (a) global shutter sampling mechanism - each row of a frame is exposed at the same time instance (b) rolling shutter sampling mechanism - each row of a frame is exposed at a different time instance \textcolor{black}{\cite{Gu2010}. $\Delta t_1$, $\Delta t_2$ and $\Delta t_3$ respectively denote reset, exposure, and readout time periods for one row.}}
  \label{Fig: Global Shutter vs Rolling Shutter}
\end{figure}

In this work, the phenomenon introduced in \cite{su2014} is further studied and an analytical model is derived \textcolor{black}{to explore how the frequency of mains-powered illumination, and so ENF} is shifted and is attenuated in relation to idle period length. Among the new \textcolor{black}{ENF components} caused by a particular idle period, the two with the most power are explored. Model predictions are verified by simulation results. Next, based on the model developed, a novel \textcolor{black}{idle period estimation approach} is proposed. To the best of our knowledge, the only work \textcolor{black}{utilising ENF for camera forensics} is that of Hajj-Ahmad et al. \cite{Hajj-Ahmad2016-ENF}. They {mainly rely on computing} the time needed to read one frame based on vertical phase analysis \textcolor{black}{\cite{Hajj-Ahmad2015}}, {i.e.\textcolor{black}{,} estimation of ENF phase for each row}. For this purpose, the sequence of the mean luminance values of the $i^{th}$ rows over each frame is  treated as a separate time series, leading to a sampling rate that equals the \textit{video frame rate}. This  does not satisfy the Nyquist Criteria \cite{Proakis-DigitalSignalProcessingPrinciplesAlgorithmsAndApplications}. Hence this approach \textcolor{black}{depends on} alias ENF and it is a challenge to apply this technique to videos where the frame rate is  a divisor of nominal ENF, e.g.\textcolor{black}{,} a  frame rate of 25 fps with nominal ENF at 50 Hz in EU or of 30 fps at 60 Hz nominal ENF in the US. This is because alias ENF is obtained at 0 Hz DC component in such a condition. It may also be a challenge for the vertical phase analysis approach to operate in a noisy video, where the power of \textcolor{black}{the captured} ENF is weak for a reliable phase estimation.

Our model and analysis also leads to a novel time-of-recording verification technique that performs better than the current techniques \cite{MinWu-SeeingENFPower-signature-basedtimestampfordigitalmultimediaviaopticalsensingandsignalprocessing}, \cite{su2014}, \cite{patent:20150356992}. A systematic search of possible ENF \textcolor{black}{components} that emerge as a result of the idle period, followed by idle period assumptions in each \textcolor{black}{component} and interpolation of missing samples for each assumption result in better quality ENF signal estimations. A better quality of ENF signal consequently leads to a better time-of-recording verification performance. In summary, the main contributions of this work include: (1) a model for where \textcolor{black}{frequency of main ENF harmonic} is shifted and how the power of \textcolor{black}{the captured} ENF is attenuated, depending on idle period length; (2) a novel {\textcolor{black}{idle period estimation technique targeted at camera forensics}; (3) a novel time-of-recording verification method for videos captured using a rolling shutter mechanism.

\begin{figure}[t]
\centering
\includegraphics[width=65mm]{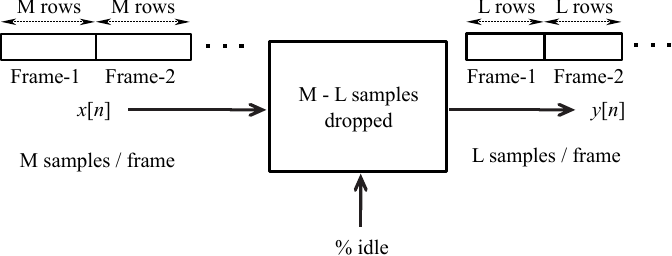}
%\vspace{2mm}
\caption{\textcolor{black}{Inherent reduction in captured luminance samples in a video due to the implementation of idle period at the end of each frame by camera: $M$ is the number of luminance samples, i.e., one sample per row, that the rolling shutter mechanism is able to capture during one frame period presuming no idle period exists. $M-L$ is the number of samples discarded due to the idle period, and hence $L$ is the actual number of samples, and so rows, per frame. $x[n]$ is the time-series of the illumination samples that can be obtained over the video duration, if there is no idle period. $y[n]$ is the actual time-series of the illumination samples as a result of abandoning the samples in the idle period of each frame.}}
\label{IdlePeriod_SampleReduction}
%\vspace{2mm}
\end{figure}

\section{The Impact of Rolling Shutter on ENF}
\label{The Impact of Rolling Shutter}
%In this section,  a comparison between global shutter and rolling shutter mechanisms is provided so as to provide key insights into the differences in ENF signal computation for the two different sampling mechanisms. Second, the phenomenon that the "ENF harmonic is shifted to some other frequency due to idle period",  as discussed in \cite{su2014} is further investigated. Accordingly, an analytical model providing how and where ENF harmonic is shifted and is attenuated depending on idle period length is developed. The variation of the most powerful ENF harmonic depending on the idle period length is illustrated. 

%\subsection{Global Shutter vs. Rolling Shutter} \label{Global shutter vs. Rolling Shutter}
In the global shutter mechanism, widely used in CCD sensors, an entire frame is exposed at one time instance, i.e.\textcolor{black}{,} each row of pixels of a single frame is sampled simultaneously. Whereas in rolling shutter mechanism, mostly used in CMOS sensors, each row of pixels in a frame is captured sequentially at different time instances. Fig. \ref{Fig: Global Shutter vs Rolling Shutter} illustrates the timing for the two distinct procedures for a single frame \cite{Gu2010}. The state-of-the-art ENF estimation techniques in the literature are basically designed by taking these sampling mechanisms into consideration. The approach in \cite{Garg2011} mainly \textcolor{black}{relies on} the global shutter phenomenon. In this method, one illumination sample is obtained per frame by averaging all steady pixels within the frame. Hence, the sampling frequency in this technique is essentially the  \textit{video frame rate}, which is much lower than \textcolor{black}{twice} the nominal ENF frequency. As this does not satisfy the Nyquist criteria \cite{Proakis-DigitalSignalProcessingPrinciplesAlgorithmsAndApplications}, it has to work with alias ENF. However when the video frame rate is  a divisor  of the nominal ENF, alias ENF is observed at the 0 Hz DC component. So, it is a great challenge for this scheme to work under this condition. The super-pixel based ENF estimation technique in \cite{Vatansever2017_SPL} is more generic and is able to operate in different sampling mechanisms. However, as in \cite{Garg2011}, this technique also depends on alias ENF since one illumination sample only is obtained per frame by steady super-pixels average within the frame.

The method in \cite{su2014} and \cite{patent:20150356992} is based on the rolling shutter phenomenon. It leads to a sampling frequency as high as \textit{number of rows $\times$ camera frame rate}, as each row is treated as a different illumination sample. Consequently, since the number of samples is much higher than \textcolor{black}{twice} the ENF frequency (Nyquist rate), alias ENF is not a phenomenon for this scheme.

%\begin{figure}[t]
%\centering
%\includegraphics[width=80mm]{RollingShutterMechanism}
%%\vspace{2mm}
%\caption{Rolling shutter Mechanism: an idle period is applied after sampling of the rows successively in each frame}
%\label{RollingShutter}
%%\vspace{2mm}
%\end{figure}
%%\begin{figure}[thp]

\begin{figure}[t]
\centering
\includegraphics[width=76mm]{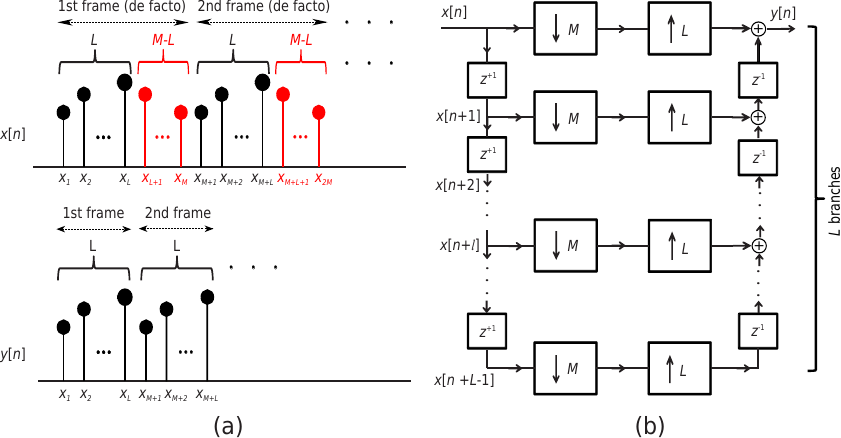}
%\vspace{2mm}
\caption{\textcolor{black}{Illustration of Rolling Shutter sampling mechanism \cite{su2014} (a) in time domain; (b) based on poly-phase decomposition.}}
\label{Fig_FilterBank}
%\vspace{2mm}
\end{figure}
%\begin{figure}[thp]



\begin{figure}[t]
\centering
\includegraphics[width=88mm]{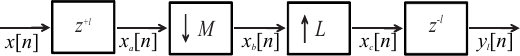}
%\vspace{2mm}
\caption{\textcolor{black}{The stages at the \textcolor{black}{$l_{th}$} branch of the poly-phase model in Fig. \ref{Fig_FilterBank} (b).}}
\label{PolyPhaseOneLine}
%\vspace{2mm}
\end{figure}

\begin{figure}[t]
\centering
\includegraphics[width=88mm]{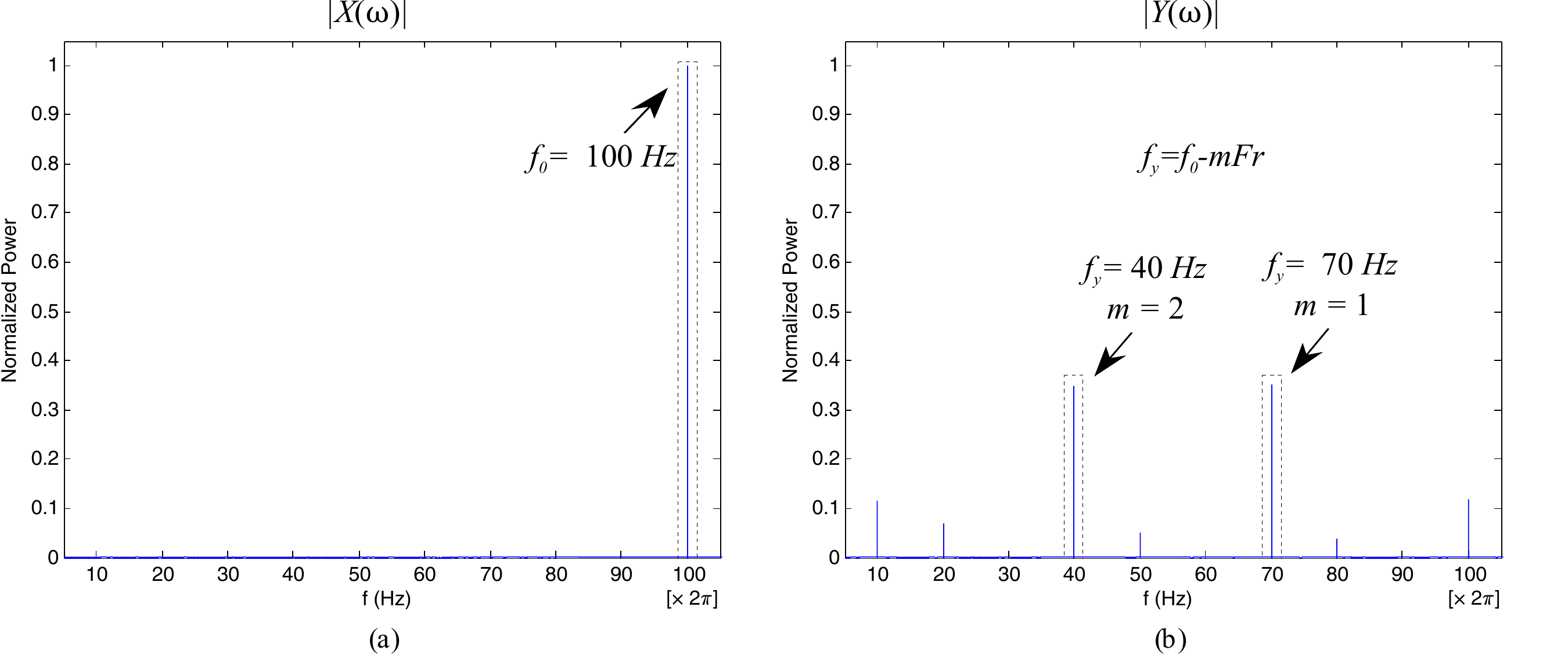}
%\vspace{2mm}
\caption{\textcolor{black}{Frequency spectrum of a video with 30 fps captured in EU, where nominal ENF is 50 Hz, provided that (a) no idle period is applied by the camera: the only ENF component appears at 100 Hz frequency (main ENF harmonic) (b) $45\%$ idle period is applied: new ENF components emerge in some other frequencies. The power of the captured ENF noticeably reduces due to the idle period.}}

\label{IdlePeriod_SampleOutput}
%\vspace{2mm}
\end{figure}

\begin{figure*}[t]
  \centering
  \subfigure[]{\includegraphics[width=67mm]{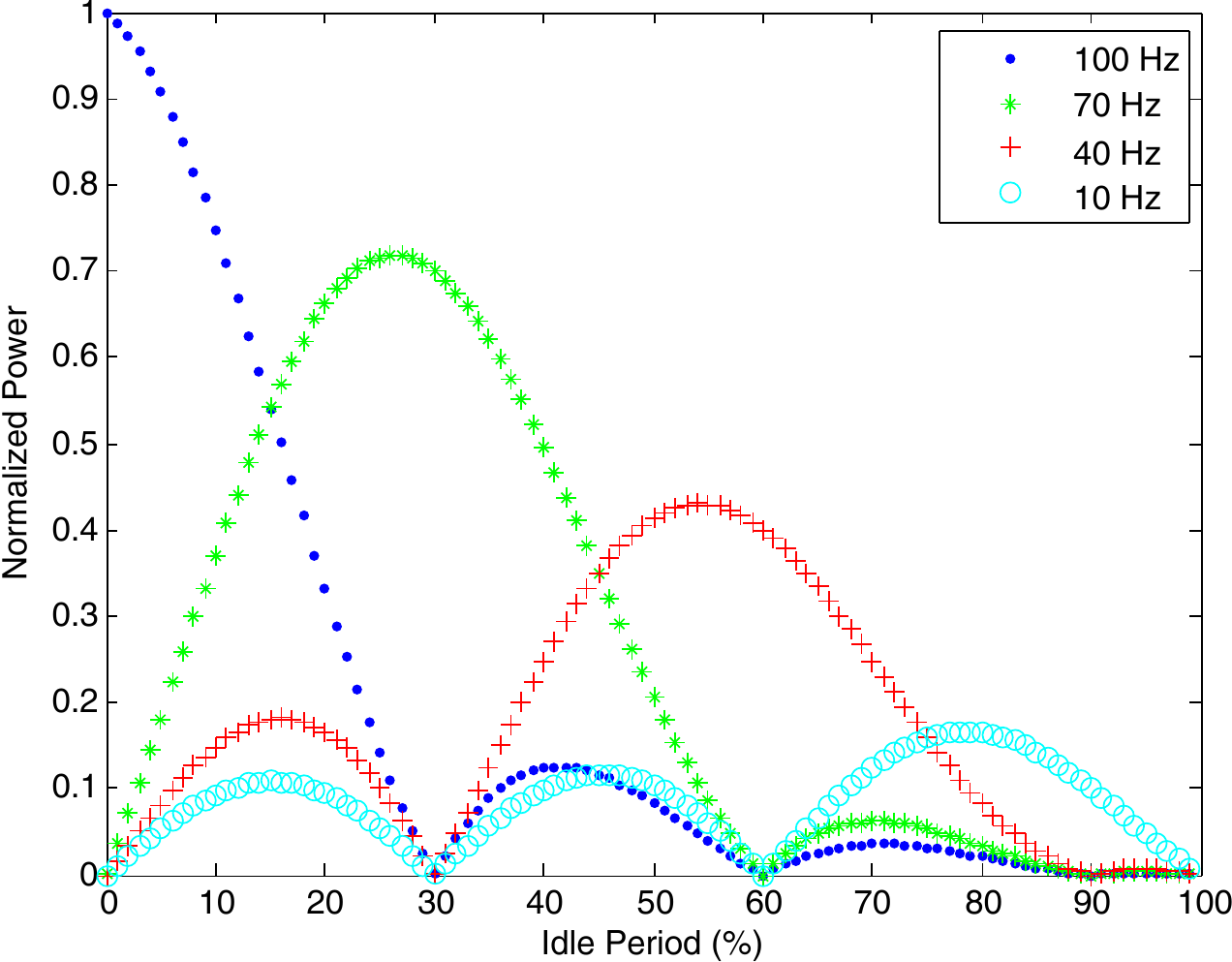}}
  %\hfill
  %\enskip
  %\qquad
  \hskip 6em
  \subfigure[]{\includegraphics[width=67mm]{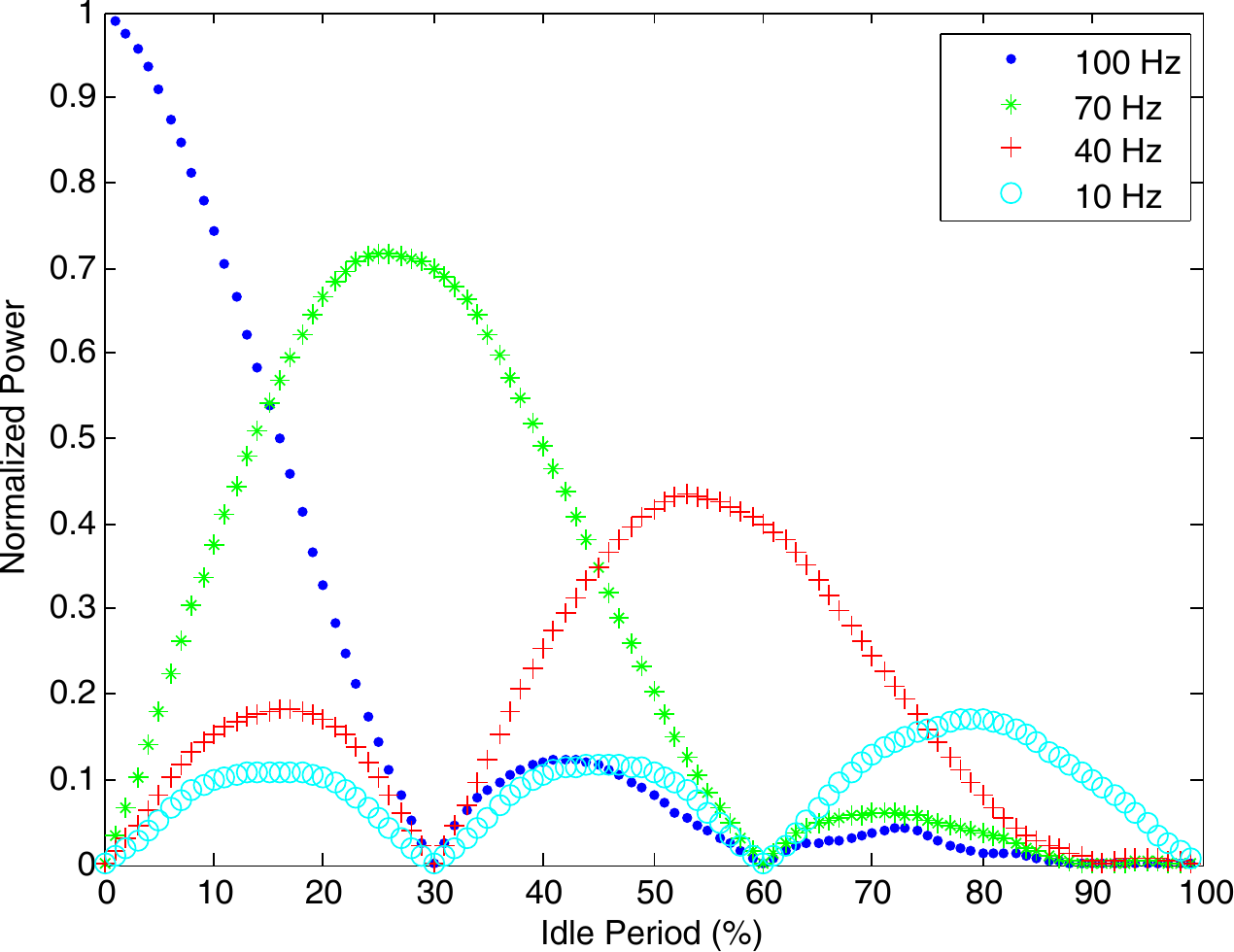}}
  \caption{\textcolor{black}{Variation in frequency of main ENF harmonic} vs. idle period for a video with 30 fps captured in EU. (a) The proposed analytical model, (b) Simulation\textcolor{black}{.}}
  \label{idle_vs_ENF_for_30fp_50Hz_Model&Sim}
\end{figure*}

\subsection{Idle Period Effect: \textcolor{black}{Variation in Frequency of Captured ENF Component}}
%Although, the rolling shutter mechanism leads to a high sampling frequency and consequently to avoid alias frequency, it brings with it the idle period issue in each frame. That is, some illumination samples in each frame are not captured. Due to these missing samples in each frame, the ENF harmonic is shifted to some other frequency \cite{su2014}. In this subsection, by following Su et al.'s work \cite{su2014}, this phenomenon is further studied and the gaps are filled. Accordingly, an analytical model is built illustrating how and where ENF harmonic is moved and is weakened in relation to idle period length.

%\section{Rolling Shutter Effect on ENF based Video Forensics}
%
%\subsection{Model and Proposed Approach}
%\subsubsection{Filter Bank Model}

%\subsubsection{Filter Bank Illustration}

Although the rolling shutter mechanism can lead to an increase of sampling frequency and consequently  avoids alias frequency, it brings with it the idle period issue. That is, some illumination samples in each frame are \textcolor{black}{missed \cite{su2014}. A representation of this phenomenon is depicted in Fig. \ref{IdlePeriod_SampleReduction}.} Here, $M$ represents the number of samples, i.e.\textcolor{black}{,} the number of rows that the rolling shutter mechanism is capable of capturing during one frame period. That is, the sampling mechanism has a capacity of sampling $M$ rows over a frame period assuming no sample is lost, i.e.\textcolor{black}{,} if the idle time is zero seconds. $M-L$ denotes the number of samples that are discarded during the idle period, so $L$ is the number of remaining samples, i.e.\textcolor{black}{,} \textcolor{black}{the actual number of rows per frame} \textcolor{black}{($L\leq$ M)}. $x[n]$ is the time-series of the illumination samples, if there is no idle period. $y[n]$ is the output time-series of the remaining illumination samples after dropping the samples in the idle period of each frame.

\textcolor{black}{A time-domain representation of the above procedure is depicted in Fig. \ref{Fig_FilterBank} (a) \cite{su2014}, and a realization of poly-phase decomposition \cite{Book_MultirateSignal} of this representation is illustrated in Fig. \ref{Fig_FilterBank} (b) \cite{su2014}}. In this model, for simplicity, the anti-aliasing filter is omitted considering the narrow band nature of ENF. In each branch of the model, the input signal is shifted back in time, \textcolor{black}{Eq. \ref{Eq:(1)}}, followed by an M-fold down-sampling filter, \textcolor{black}{Eq. \ref{Eq:(2)}}. Next, an L-fold up-sampling filter is applied, \textcolor{black}{Eq. \ref{Eq:(3)}}. Then the signal is shifted forward in the appropriate order, \textcolor{black}{Eq. \ref{Eq:(4)}}. \textcolor{black}{For the \textcolor{black}{$l_{th}$} branch, these stages are demonstrated in Fig. \ref{PolyPhaseOneLine}. Accordingly, DTFT (\textcolor{black}{Discrete Time Fourier Transform}) of $x_a[n]$, $x_b[n]$, $x_c[n]$ and $y_l[n]$ are respectively obtained as follows:}

\textcolor{black}{
\begin{align}
X_a \left( e^{ j \omega} \right) = X \left( e^{ j \omega} \right) e^ { j \omega l }
\label{Eq:(1)}
\end{align}
}
\textcolor{black}{
\begin{align}
\begin{split}
X_b \left( e^{ j \omega} \right) &= \frac{1}{M} \sum_{m=0}^{M-1} X_a \left( e^{ j \left( \frac{\omega}{M} - \frac{ 2\pi m}{M} \right) } \right) \\
		&= \frac{1}{M} \sum_{m=0}^{M-1} \left[ X \left( e^{ j \left( \frac{\omega}{M} - \frac{ 2\pi m}{M} \right) } \right) e^ { j \left( \frac{\omega}{M} - \frac{ 2\pi m}{M} \right) l } \right] \\
\end{split}
\label{Eq:(2)}
\end{align}
}
\textcolor{black}{
\begin{align}
\begin{split}
X_c \left( e^{ j \omega} \right) &= X_b \left( e^{ j \omega L } \right) \\
		&= \frac{1}{M} \sum_{m=0}^{M-1} \left[ X \left( e^{ j \left( \frac{ \omega L }{M} - \frac{ 2\pi m}{M} \right) } \right) e^ { j \left( \frac{ \omega L }{M} - \frac{ 2\pi m}{M} \right) l } \right] \\
\end{split}
\label{Eq:(3)}
\end{align}
}
\textcolor{black}{
\begin{align}
Y_l \left( e^{ j \omega} \right) = X_c \left( e^{ j \omega } \right) e^{ - j \omega l }
\label{Eq:(4)}
\end{align}
Then, DTFT of the signal at the end of the $l^{th}$ branch, $Y_l \left( e^{ j \omega} \right)$ can be expressed as follows:
\begin{align}
\label{Eq:one line}
Y_l \left( \omega \right) = \frac{1}{M} \left( \sum_{m=0}^{M-1} X \left( \frac{\omega L-2\pi m}{M} \right) e^{ j \frac{\omega L-2\pi m}{M} l} \right) e^{-j \omega l}
\end{align}
}
%
%\begin{align}
%\frac{1}{M} \left( \sum_{m=0}^{M-1} X \left( \frac{\omega L+2\pi m}{M} \right) e^{j \frac{\omega L+2\pi m}{M} l} \right) e^{-j \omega l}
%\end{align}
%
\textcolor{black}{It should be noted in Eq. \ref{Eq:one line} that $Y_l \left( e^{ j \omega} \right)$ is shown in the form $Y_l \left( \omega \right)$ for simplicity. From this point on, $Y \left( e^{ j \omega} \right)$, $X \left( e^{ j \omega} \right)$ and other relevant frequency domain variables are shown in similar format. Consequently, when all the signals at the end of each branch of the poly-phase model are combined together, the output signal $Y(\omega)$ is formed as follows \cite{su2014}:}
\textcolor{black}{
\begin{align}
\begin{split}
Y(\omega) &= \sum_{l=0}^{L-1} Y_l(\omega) \\
		&= \sum_{l=0}^{L-1} \frac{1}{M} \left( \sum_{m=0}^{M-1} X \left( \frac{\omega L-2\pi m}{M} \right) e^{j \frac{\omega L-2\pi m}{M} l} \right) e^{-j \omega l} \\
		&= \sum_{m=0}^{M-1} X \left( \frac{\omega L-2\pi m}{M} \right) F_m(\omega) \\
\end{split}
\label{Eq_Filterbank}
\end{align}
where
\begin{align}
F_m (\omega)=\frac{1}{M} \sum_{l=0}^{L-1} e^{-j \frac{\omega (M-L)+2 \pi m}{M} l}
\label{Fm}
\end{align}
}
%\begin{figure*}[t]
%  \centering
%  \subfigure[]{\includegraphics[width=80mm]{30fps_480P}}
%  %\hfill
%  \quad
%  \subfigure[]{\includegraphics[width=80mm]{30fps_480P_60Hz}}
%  \caption{The proposed analytical model - variation of ENF Harmonic vs. duration of idle period for a video with 30 fps (a) for EU, where nominal ENF is 50 Hz, (b) for US, where nominal ENF is 60 Hz)}
%  \label{idle_vs_ENF_for_30fp_50Hz_60Hz}
%\end{figure*}
%
%\begin{figure*}[t]
%  \centering
%  \subfigure[]{\includegraphics[width=80mm]{30fps_480P_simulation}}
%  %\hfill
%  \quad
%  \subfigure[]{\includegraphics[width=79mm]{30fps_480P_60Hz_simulation}}
%  \caption{Simulation - variation of ENF Harmonic vs. duration of idle period for a video with 30 fps (a) for EU, where nominal ENF is 50 Hz, (b) for US, where nominal ENF is 60 Hz)}
%  \label{idle_vs_ENF_for_30fp_50Hz_60Hz_sim}
%\end{figure*}

\subsection{Proposed Model}
\label{AnalyticalModel}
\textcolor{black}{In Eq. \ref{Eq_Filterbank}, $F_m$ specifies the amount of attenuation in the ENF signal depending on the proportion of $L$ to $M$. Accordingly, by \textcolor{black}{disregarding} $F_m$, the $\omega_y$ value making $\abs{X \left( \frac{\omega_y L-2\pi m}{M} \right)}$  the same as $\abs{X(\omega_0)}$ is the shifted angular frequency for a specific idle period, where $\omega_0$ is nominal angular electric frequency. To find $\omega_y$ value, the following cases should be analysed:}

%Denoting $X \left( \frac{\omega L-2\pi m}{M} \right)$ as $\hat{X}(\omega)$, 

%Accordingly, considering the fact that Fourier transform is an even function, the following equivalents can be obtained:}
%\begin{align}
%%X \left( \frac{\omega L+2\pi m}{M} \right) \approxeq A \delta (\omega-\omega_0)
%\abs{X(\omega_0)} = \abs{X(-\omega_0)} = \abs{Y(\omega_y)} = \abs{Y(-\omega_y)}
%\label{Eq_Filterbank_Case1}
%\end{align}
\textcolor{black}{
\noindent \textbf{Case 1}:
\begin{align}
\abs{X(\omega_0)} =  \abs{X \left( \frac{\omega_y L - 2\pi m}{M} \right)}
\label{Eq_Filterbank_Case1_abs}
\end{align}
%\begin{align}
%%\begin{split}
%\abs{\hat{X}(\omega_y)} = \abs{X \left( \frac{\omega_y L - 2\pi m}{M} \right)} \approxeq A \delta (\omega-\omega_0)
%%\abs{X(\omega_0)} &=  \abs{Y(\omega_y)} \\
%%&= Y \left( \frac{\omega_y L - 2\pi m}{M} \right) \\
%%\end{split}
%\label{Eq_Filterbank_Case1_abs}
%\end{align}
%where $\delta$ is delta function, and $A$ is a positive number, i.e. magnitude.
Accordingly;
\begin{align}
%X \left( \frac{\omega L+2\pi m}{M} \right) \approxeq A \delta (\omega-\omega_0)
\omega_0 = \frac{\omega_y L - 2\pi m}{M}
\label{Eq_Filterbank_Case1}
\end{align}
$\omega_y$ and $\omega_0$ can be expressed as follows: 
\begin{align}
\omega_y=\frac{2\pi f_y}{F_rL} \quad \text{and} \quad \omega_0=\frac{2\pi f_0}{F_rM}
\label{Eq_omega}
\end{align}
where $f_0$ is nominal illumination frequency (\textcolor{black}{twice} the ENF), $f_y$ is the emerging shifted illumination frequency, and $F_r$ is video frame rate.
By substituting $\omega$ and $\omega_0$ equivalents, Eq. \ref{Eq_Filterbank_Case1} can be rewritten as follows:
\begin{align}
\frac{2 \pi f_0}{F_rM} = \frac{2 \pi f_y}{F_rL} \frac{L}{M} - \frac{2\pi m}{M} \frac{F_r}{F_r}
\label{Eq_Case1_eq}
\end{align}
From the Eq. \ref{Eq_Case1_eq}, for \textbf{Case 1}, $f_y$ can be obtained as:
\begin{align}
f_y = f_0 + mF_r
\label{Eq_Case1_f}
\end{align}
where, $f_y<\frac{F_rL}{2}$, which comes from the Nyquist theorem considering $F_rL$ is the sampling frequency. Accordingly, from the Eq. \ref{Eq_Case1_f}, the range of $m$ for \textbf{Case 1} can be obtained as follows:
\begin{align}
m < \frac{L}{2} - \frac{f_0}{F_r}, \quad m \in \mathbb{W}
\label{Eq_S1}
\end{align}
%Then, the set of m values for \textbf{Case 1} are obtained as follows:
%\begin{align}
%m \in{S_1}=\left\lbrace 0,1,..., \frac{L}{2} - \ceil*{\frac{f_0}{F_r} } \right\rbrace
%\label{Eq_S1}
%\end{align}
}
\textcolor{black}{
\noindent \textbf{Case 2}:
Denoting $X \left( \frac{\omega L-2\pi m}{M} \right)$ as $\hat{X}(\omega)$, the following equivalents can be obtained based on the phenomenon that Fourier transform is an even function:
\begin{align}
%X \left( \frac{\omega L+2\pi m}{M} \right) \approxeq A \delta (\omega-(2\pi-\omega_0))
\abs{\hat{X}(\omega_y)} =  \abs{\hat{X}(-\omega_y)} = \abs{X \left( \frac{-\omega_y L - 2\pi m}{M} \right)}
\label{Eq_Filterbank_Case2_abs1}
\end{align}
Similarly,
\begin{align}
\abs{X(\omega_0)} = \abs{X(-\omega_0)}
\label{Eq_Filterbank_Case2_abs1}
\end{align}
Then, the following expressions can be formed:
\begin{align}
\abs{X(-\omega_0)} = \abs{X \left( \frac{-\omega_y L - 2\pi m}{M} \right)}
\label{Eq_Filterbank_Case2_abs1}
\end{align}
Accordingly, 
\begin{align}
\omega_0 = \frac{\omega_y L + 2\pi m}{M}
\label{Eq_Filterbank_Case2}
\end{align}
By substituting $\omega$ and $\omega_0$ equivalents in Eq. \ref{Eq_omega}, Eq. \ref{Eq_Filterbank_Case2} can be rewritten as follows:
\begin{align}
\frac{2 \pi f_0}{F_rM} = \frac{2 \pi f_y}{F_rL} \frac{L}{M} + \frac{2\pi m}{M} \frac{F_r}{F_r}
\label{Eq_Case2_eq}
\end{align}
From the Eq. \ref{Eq_Case2_eq}, for \textbf{Case 2}, $f_y$ results as:
\begin{align}
f_y = f_0 - mF_r
\label{Eq_Case2_f}
\end{align}
From the Eq. \ref{Eq_Case2_f}, the set of $m$ values for \textbf{Case 2}, for $f_y > 0$, yield the follows:
\begin{align}
%m \in{S_2}=\left\lbrace 0,1,..., \floor*{\frac{f_0}{F_r} } \right\rbrace
m < \frac{f_0}{F_r}, \quad m \in \mathbb{W}
\label{Eq_S2}
\end{align}
It should be noted that the set of $m$ values in Eq. \ref{Eq_S2} are for the $f_y$ expression in Eq. \ref{Eq_Case2_f} only. Similarly, the set of $m$ values provided in Eq. \ref{Eq_S1} are for the $f_y$ expression in Eq. \ref{Eq_Case1_f} only. Accordingly, the following pair of equivalences are obtained:
}

\begin{align}
  f_y=\begin{cases}
    f_0+mF_r, & m < \frac{L}{2} - \frac{f_0}{F_r}.\\
    f_0 - mF_r, & m < \frac{f_0}{F_r}.
  \end{cases}
  \label{Shifted_ENF}
\end{align}
where $0<f_y<\frac{LF_r}{2}$. \textcolor{black}{Any of the ENF components of $f_y$ having relatively high SNR (Signal-to-Noise Ratio) can be used to estimate the ENF variations along time.}
\textcolor{black}{
By using \textcolor{black}{the} corresponding pair of equivalences \textcolor{black}{in Eq. \ref{Shifted_ENF}}, the $m$ value making $\abs {Y(\omega)}$ maximum can consequently be obtained as follows:
%To check if there is any common $m$ value for Eq. \ref{Eq_Case1_f} and Eq. \ref{Eq_Case2_f}, the following analysis can be performed:
%\begin{align}
%\begin{split}
%f_0 + m F_r & \stackrel{?}{=} f_0 - m F_r \\
%m F_r & \stackrel{?}{=} - m F_r
%\end{split}
%\label{Eq_Case1_2_inequality}
%\end{align}
%Accordingly, it is not possible for the expression to be valid. Therefore different values of $m$ result in different $f$ values.
\begin{align}
\begin{split}
m_p & = \argmax_{m} \abs {X_m \left(\frac{\omega L+2\pi m}{M} \right) \times F_m(\omega)}
%	&=\argmax_{m \in S_1 \cup S_2 } {F_m (\omega_m)}
\end{split}
\label{Eq_arg_max}
\end{align}
%Here, in Eq. \ref{Eq_arg_max}, since the $X_m$ can be considered as a delta function, referring to Eq. \ref{Eq_Filterbank_Case1} and Eq. \ref{Eq_Filterbank_Case2}, it takes $0$ value for the $\omega$ values other than $\omega_m$.
The corresponding $f_{y}$ for the resulting $m_p$, which can be computed from the Eq. \ref{Eq_Case1_f} or Eq. \ref{Eq_Case2_f}, is the strongest emerging \textcolor{black}{ENF component} for a particular idle period, i.e.\textcolor{black}{,} $\frac{M-L}{M} \times 100$.} \textcolor{black}{Fig. \ref{IdlePeriod_SampleOutput} (a) illustrates frequency spectrum for a video with 30 fps captured in EU, where nominal ENF is 50 Hz, provided that no idle period is applied by the camera. As can be seen from the figure, the only ENF component appears at 100 Hz frequency (main ENF harmonic). Fig. \ref{IdlePeriod_SampleOutput} (b) provides emerging ENF components if $45\%$ idle period is applied by the camera. Here, the strongest ENF component appears at 70 Hz frequency ($m = 1$), and the second strongest one appears at 40 Hz ($m = 2$). It can be highlighted that the power of 40 Hz and 70 Hz ENF components are very close since this idle period is a transition point for the greatest ENF components. Fig. \ref{IdlePeriod_SampleOutput} (b) also depicts that the power of the captured ENF noticeably reduces due to idle period. Here, the power is normalized. It is also notable that the greater ENF harmonics and emerging ENF components are disregarded.}

Fig. \ref{idle_vs_ENF_for_30fp_50Hz_Model&Sim} (a) illustrates the variation \textcolor{black}{in frequency of the strongest ENF component} depending on the proportion of the idle period length (in $\%$) for a video with 30 fps captured in EU (50 Hz mains power grid). Accordingly, the strongest ENF \textcolor{black}{component}, 100 Hz for illumination, is replaced with 70 Hz, 40 Hz and 10 Hz respectively for 15\%, 45\%, 75\% idle periods. It can also be noticed from the figure that, the power of \textcolor{black}{the captured} ENF  decreases as the idle period increases. These outcomes are also validated by simulation results. As can be seen from Fig. \ref{idle_vs_ENF_for_30fp_50Hz_Model&Sim} (b), the simulation findings are almost the same as those obtained via the proposed model. The slight differences are most likely caused by additional \textcolor{black}{ENF components} emerging due to the effect of the 2nd illumination harmonic, 200 Hz, which \textcolor{black}{comes} from the absolute cosine form of the illumination signal. Although there may be such extra \textcolor{black}{ENF components}, it is validated from both the simulation and the proposed model results that the strongest two \textcolor{black}{ENF components} at any idle period emerge from the 1st illumination harmonic.

%\textcolor{black}{The developed model is validated with simulation results. They exactly match.}
%
%\textcolor{black}{The derived model is also verified with the simulation results.}

%\textcolor{black}{In order to analyze the effect of idle period duration in ENF estimation, CMOS sensor image acquisition process is simulated. Basically, from a ground truth ENF signal, illumination signal is obtained and this signal is sampled as in rolling shutter mechanism.}

%\begin{figure}[t]
%\centering
%\includegraphics[width=80mm]{ENFHarmonic_vs_idle_v3}
%%\vspace{2mm}
%\caption{Variation of ENF Harmonic versus duration of idle period}
%\label{Fig_Harmonic_vs_idle}
%%\vspace{2mm}
%\end{figure}

%Based on the Equations \ref{Eq_S1} and \ref{Eq_S2}, the set of m values can be written as below:
%
%So, for $F_r=30 fps$ (in most videos) and for $f_0=100 Hz$ (in Europe) or $f_0=120 Hz$ (in US), $f$ can be expressed as $f=100-30m$ or $f=120-30m$. For either case, the set of m values are obtained as follows:

\section{\textcolor{black}{Idle Period Estimation}}
\textcolor{black}{Camera forensics has become a significant field of research for various applications such as verification query image/video pairs can be attributed to the same source camera, and verification the source of a suspicious image/video \cite{MemonBook}, \cite{Shullani2017}.} To the best of our knowledge, the only work exploiting ENF for camera forensics is that of Hajj-Ahmad et al. \cite{Hajj-Ahmad2016-ENF}. \textcolor{black}{They basically compute camera read-out time per frame based on vertical phase analysis \textcolor{black}{\cite{Hajj-Ahmad2015}}.}
%Although this approach may be an effective tool for some cases, it has some limitations.
% it is dependent on alias ENF, so it is inapplicable for videos with frame rates that are divisors of the nominal ENF frequency.
Although the technique they introduced, i.e.\textcolor{black}{,} vertical phase analysis, may be an effective tool for some cases, it has some significant limitations. First, it is dependent on alias ENF, and so is inapplicable to videos, where the frame rate is a divisor of nominal ENF (0 Hz alias ENF). Second, it may be a challenge for their method to operate on noisy videos due to difficulty in computation of the ENF phase correctly in such conditions. In the subsections below, this state-of-the-art approach is discussed first. Then a novel \textcolor{black}{idle period estimation} approach is proposed, which can overcome the limitations of the state-of-the-art. Next, experimental work is provided for a performance evaluation.

\begin{figure*}[t]
  \centering
  \subfigure[]{\includegraphics[width=65mm]{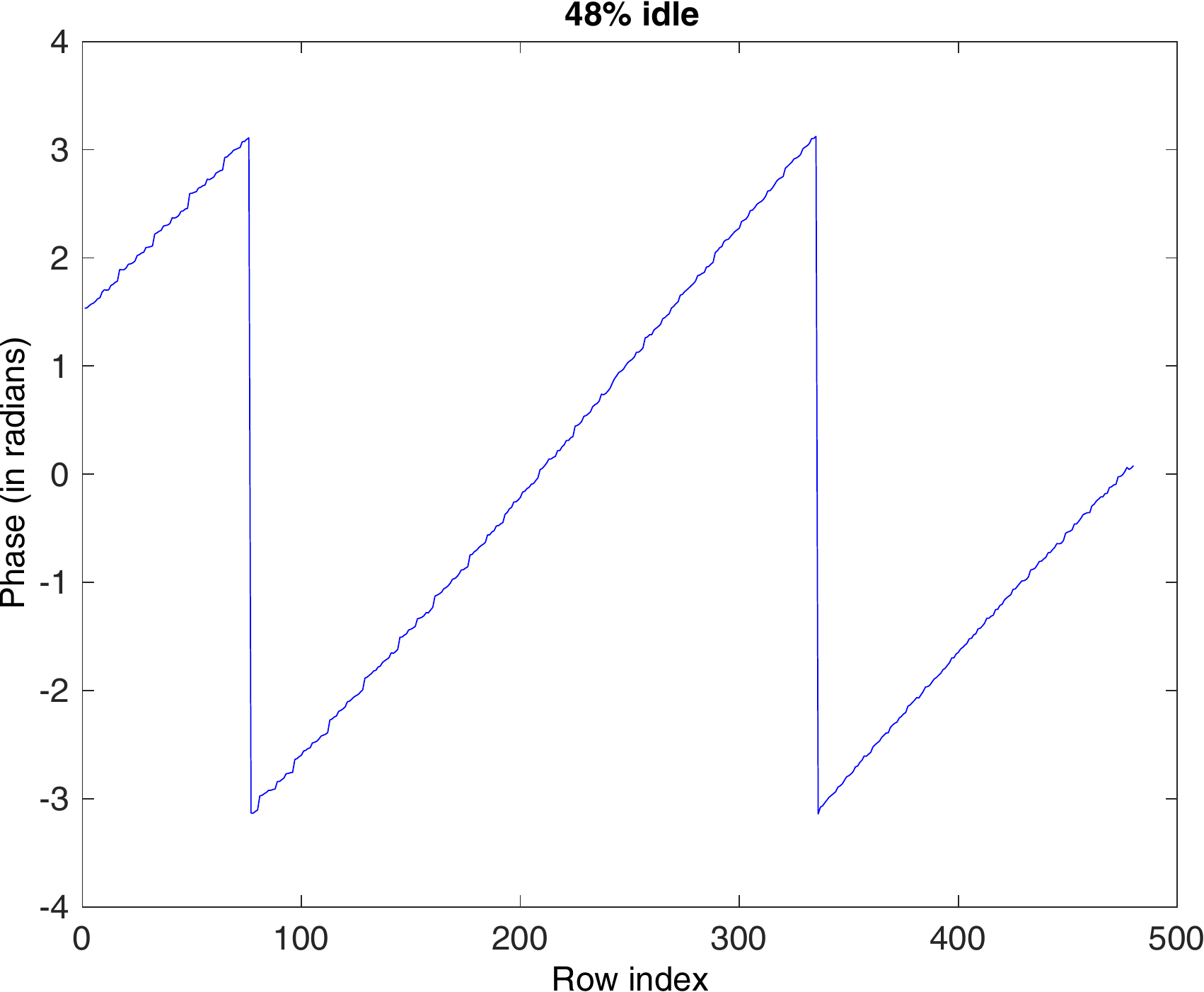}}
  %\hfill
  %\quad
  \hskip 6em
  \subfigure[]{\includegraphics[width=65mm]{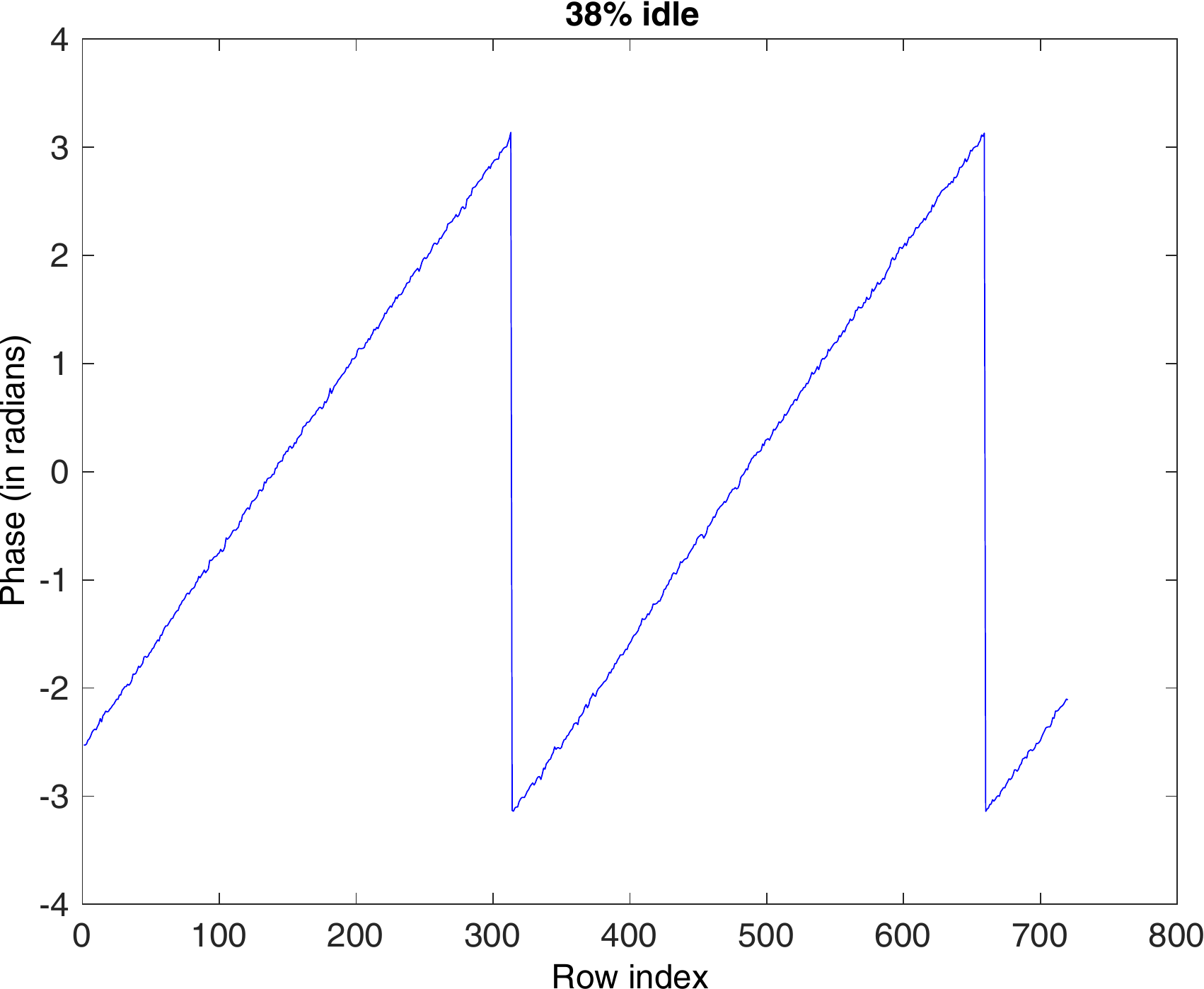}}
  \caption{(a) Vertical phase analysis for a 480 P wall-scene video, video-1, corresponding to about 48\% idle period. (b) Vertical phase analysis for a 720 P wall-scene video, video-2, corresponding to about 38\% idle period. Both videos are recorded at 30 fps by CanonPowerShot SX230 HS model camera in Turkey, where illumination frequency is 100 Hz (\textcolor{black}{twice} the nominal ENF (50 Hz))\textcolor{black}{.}}
  \label{Fig: vertical phase-1-2}
\end{figure*}

\begin{figure*}[t]
  \centering
  \subfigure[]{\includegraphics[width=65mm]{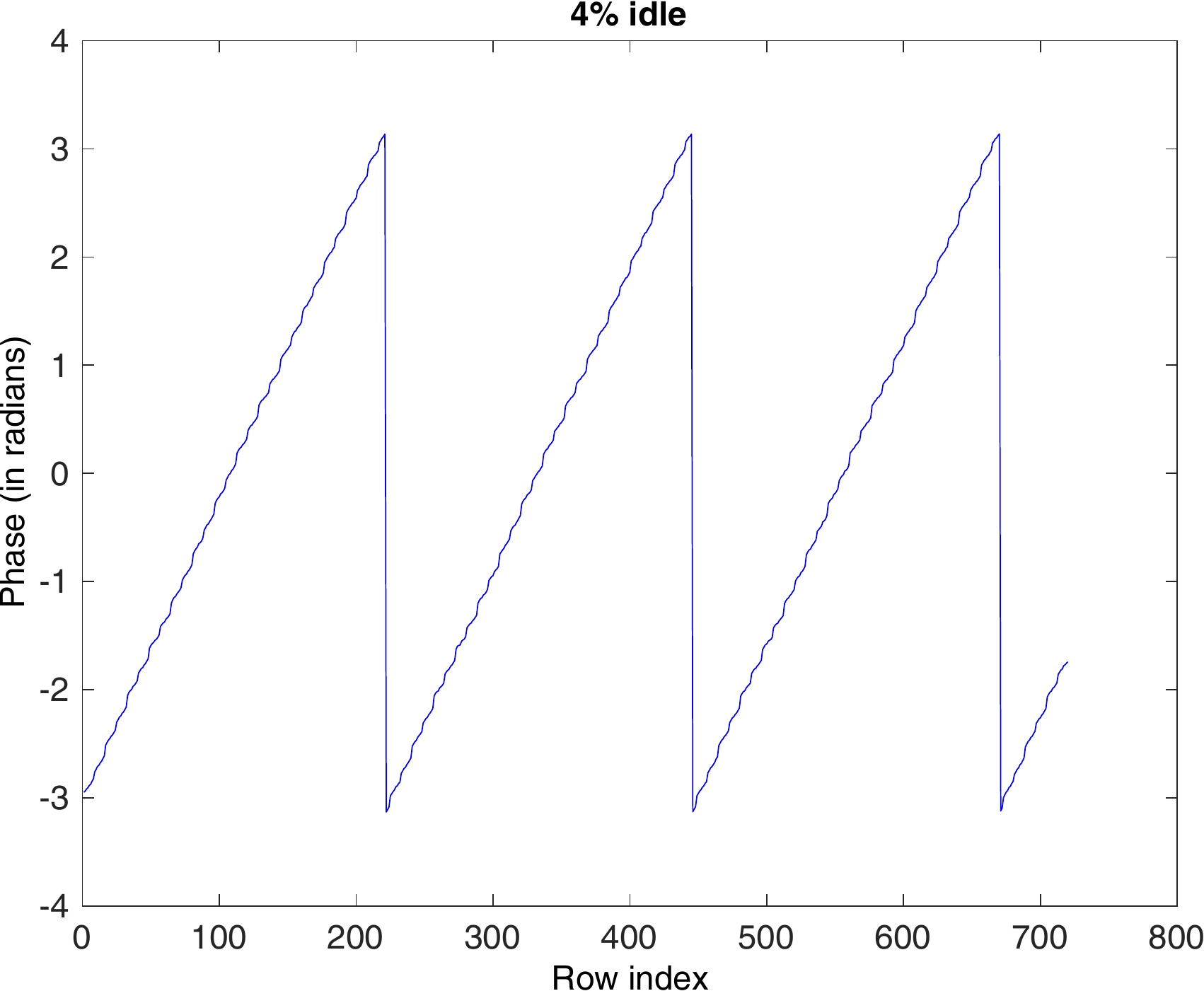}}
  %\hfill
  %\quad
  \hskip 6em
  \subfigure[]{\includegraphics[width=65mm]{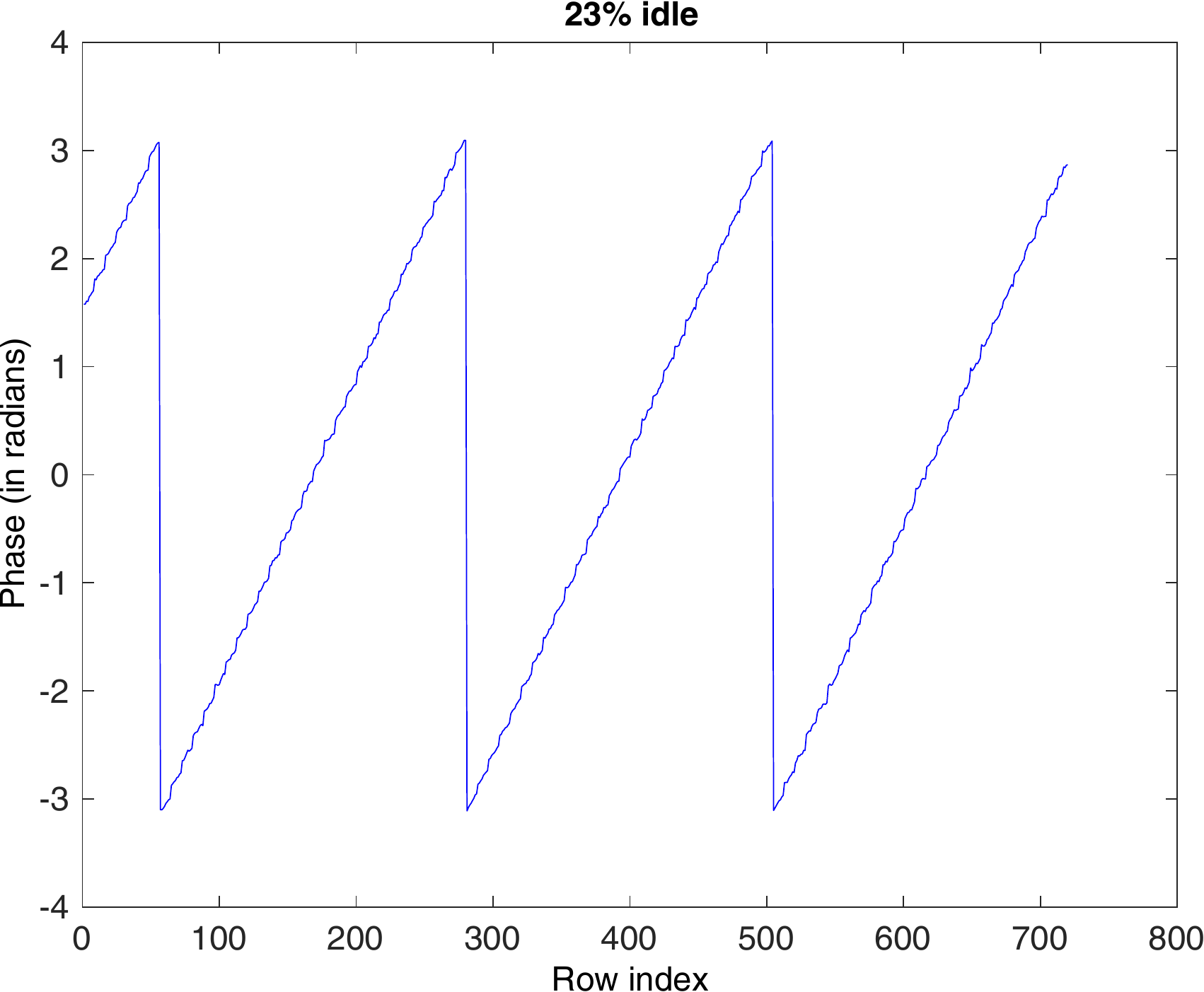}}
  \caption{(a) Vertical phase analysis for a 30 fps wall-scene video, video-3, corresponding to about 4\% idle period. (b) Vertical phase analysis for a 23.976 fps wall-scene video, video-4, corresponding to about 23\% idle period. Both videos are recorded in 720 P resolutions by Nikon D3100 model camera in Turkey, where illumination frequency is 100 Hz (\textcolor{black}{twice} the nominal ENF (50 Hz))\textcolor{black}{.}}
  \label{Fig: vertical phase-3-4}
\end{figure*}

\subsection{State-of-the-art Approach} \label{State of the Art Approach}
This subsection, first, briefly discusses how \textcolor{black}{camera read-out time can be estimated} via vertical phase analysis \cite{Hajj-Ahmad2016-ENF}, \textcolor{black}{\cite{Hajj-Ahmad2015}}. Then, an adaptation of this technique to idle period computation is presented. Results from the adapted technique, including some common properties of idle period are illustrated via experiments conducted on videos with still content (a wall-scene), which are all known to contain ENF. Then the limitations of this approach are explored.

Hajj-Ahmad et al. technique targets cameras with a  rolling shutter mechanism. It  basically computes the time needed to read one frame (frame read-out time), $T_{ro}$ by \textcolor{black}{estimating phase of ENF for each row along the video frames, i.e.\textcolor{black}{,} vertical phase analysis as} follows:
\begin{align}
T_{ro}=\frac{L \tilde{\omega}_b}{2 \pi \tilde{f}_e}
\label{Eq_read_out}
\end{align}
where $L$ is the number of rows in a frame, $\tilde{\omega}_b$ is vertical radial frequency and $\tilde{f}_e$ is the ENF component that oscillates around the nominal  frequency. $\tilde{\omega}_b$ can be computed from the slope of the vertical phase line, e.g.\textcolor{black}{,} Fig. \ref{Fig: vertical phase-1-2}, which can be obtained via the estimation of the phase of ENF, $\Phi \left[ l \right] $ ($l \in \left\lbrace 1, 2, 3, ..., L \right\rbrace$, where $L$ is number of rows) for each single row \cite{Hajj-Ahmad2016-ENF}. $\Phi \left[ l \right] $ can be extracted through the Fourier Transform of the $l$th-row time-series that can be obtained by using the $l$th-row mean intensity of each frame of the video. Since the sampling frequency of this time-series is simply the video frame rate (and hence Nyquist theorem is not satisfied), peak search in the Fourier domain is made around alias ENF. That is, the property of the increased sampling frequency (\textit{number of rows $\times$ camera frame rate}) of the rolling shutter, discussed in \ref{The Impact of Rolling Shutter}, is invalid. What this approach mainly \textcolor{black}{exploits is the ENF phase shift emerged between consecutive rows due to distinct time of sampling of each row by rolling shutter}.

Vertical phase analysis can easily be adapted to perform idle period computation. \textcolor{black}{The adapted technique mainly utilizes 2 key features, namely the number of sinusoidal cycles that would be acquired per frame if there was no idle period and and the actual number of sinusoidal cycles that is acquired per frame, i.e.\textcolor{black}{,} in the presence of idle period.} Fig. \ref{Fig: vertical phase-1-2} (a) and Fig. \ref{Fig: vertical phase-1-2} (b) illustrate the vertical phases estimated respectively for a 480 P video (video-1), and a 720 P video (video-2),  recorded by Canon PowerShot SX230HS model camera with 30 fps in Turkey, where the nominal illumination frequency is 100 Hz. From the figures, the proportion of the idle period per frame (in $\%$), $R_{T_{I}}$, can be  computed with the use of the following equation:
\begin{align}
R_{T_{I}}=100 - \frac{f_I}{F_r} \times \frac{100}{N_c}
\label{Eq_read_out}
\end{align}
where $N_c$ is the actual number of \textcolor{black}{sine cycles, i.e.\textcolor{black}{,} triangles (between $-\pi$ and $+\pi$)} obtained through vertical phase analysis. $f_I$ denotes the nominal illumination frequency, and $F_r$ is video frame rate. \textcolor{black}{$\frac{f_I}{F_r}$ results in the number of sinusoidal waves that would be acquired per frame if there was no idle period.} Accordingly $R_{T_{I}}$ values for video-1 and video-2 are obtained as 48\% and 38\% respectively. \textcolor{black}{From these findings, it can be deduced that video resolution may a factor affecting the idle period.}

Fig. \ref{Fig: vertical phase-3-4} (a) and Fig. \ref{Fig: vertical phase-3-4} (b) illustrate the vertical phases estimated respectively for a 30 fps video (video-3), and a 23.976 fps video (video-4), which are recorded by Nikon D3100 model camera in 720 P in Turkey. Accordingly, $R_{T_{I}}$ values for video-3 and video-4 are obtained as 4\% and 23\% respectively. These outcomes show that \textcolor{black}{video frame rate is another factor affecting the idle period.}

\begin{figure}[t]
\centering
\includegraphics[width=67mm]{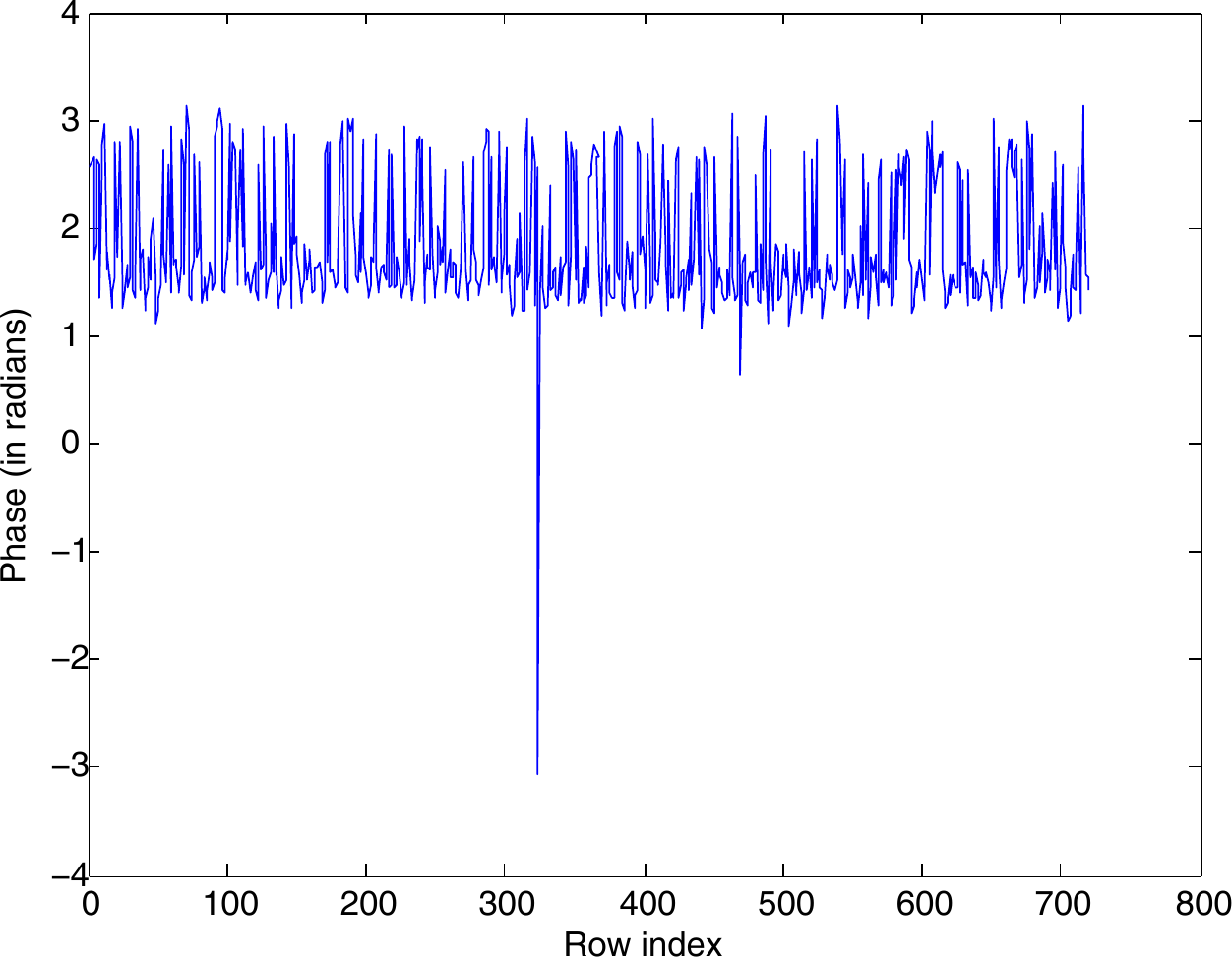}
\caption{(a) Vertical phase analysis for a 23.976 fps video, video-5 - it is a great challenge to compute the phase for this video, which is captured under illumination of \textbf{CFL} bulb by Nikon D3100 camera in Turkey\textcolor{black}{.}}
\label{Fig: vertical phase-5 and frequency domain}
\end{figure}

Although vertical phase analysis technique can provide estimations of idle period time, it has some limitations. First, when the power of \textcolor{black}{the captured} ENF in a video is weak due to various reasons including moving content, long idle period and unusual characteristic of some light sources, it may be a challenge to estimate vertical phase correctly. Such an example is illustrated in Fig. \ref{Fig: vertical phase-5 and frequency domain} (a), computed from a wall-scene (still content) video illuminated by a CFL (compact fluorescent) bulb. As can be seen from the figure, it is really tough for the phase to be estimated in such a condition. It should be noted that spectral responses of different type of illumination are different and CFL bulb illumination is one that may contribute to the quality of ENF negatively \cite{Vatansever2017}. Another limitation of the vertical phase method is that the vertical phase approach is dependent on alias ENF. Therefore, it is also a great challenge for this approach to work for videos recorded with a frame rate that is a divisor of the nominal illumination frequency. It should be recalled that alias ENF is obtained at 0 Hz in such a condition and it is extremely thorny to make an estimate of the ENF from this DC component.

%\begin{table}[!t]
%\renewcommand{\arraystretch}{1.3}
%\caption{\textcolor{black}{Idle Period Estimation Approach}}
%\label{Table_SourceCameraVerification}
%\centering
%%\begin{tabular}{p{0.5cm} p{7.35cm}}
%\begin{tabular}{p{0.25cm} p{7.35cm}}
%\hline\hline
%\centering
%\bfseries Step & \bfseries Description\\
%\hline
%
%1 & Based on the test video frame rate and nominal ENF of the connected mains power, analytical model for the variation of the ENF harmonic with the highest power, depending on idle period length, is computed using the method described in section \ref{AnalyticalModel}.\\
%
%2 & For each row, an illumination sample is computed via all the steady pixels average of the row, followed by concatenation of the estimated illumination samples of all rows of all frames, similar to the procedure in Fig. \ref{Fig_FilterBank} (a), to form the time-series for illumination variation during the video period.\\
%
%3 & Based on the derived model in Step 1, the strongest two ENF harmonics in the test video, $H_1$ and $H_2$, are located in the frequency domain (Discrete Fourier Transform) of the time-series formed.\\
%
%4 & The ratio of the power of the strongest ENF harmonic, $H_1$, to the second strongest, $H_2$, is computed. Let the ratio be $P_{H_1}/P_{H_2}$.\\
%
%5 & By using $H_1$ and $H_2$ and $P_{H_1}/P_{H_2}$, locate the corresponding range of the idle period on the analytical model illustration.\\
%
%6 & The middle of the located range is assigned as the estimated idle period.\\
%
%\hline\hline
%\end{tabular}
%\end{table}

\begin{figure*}[t]
  \centering
  \subfigure[]{\includegraphics[width=64mm]{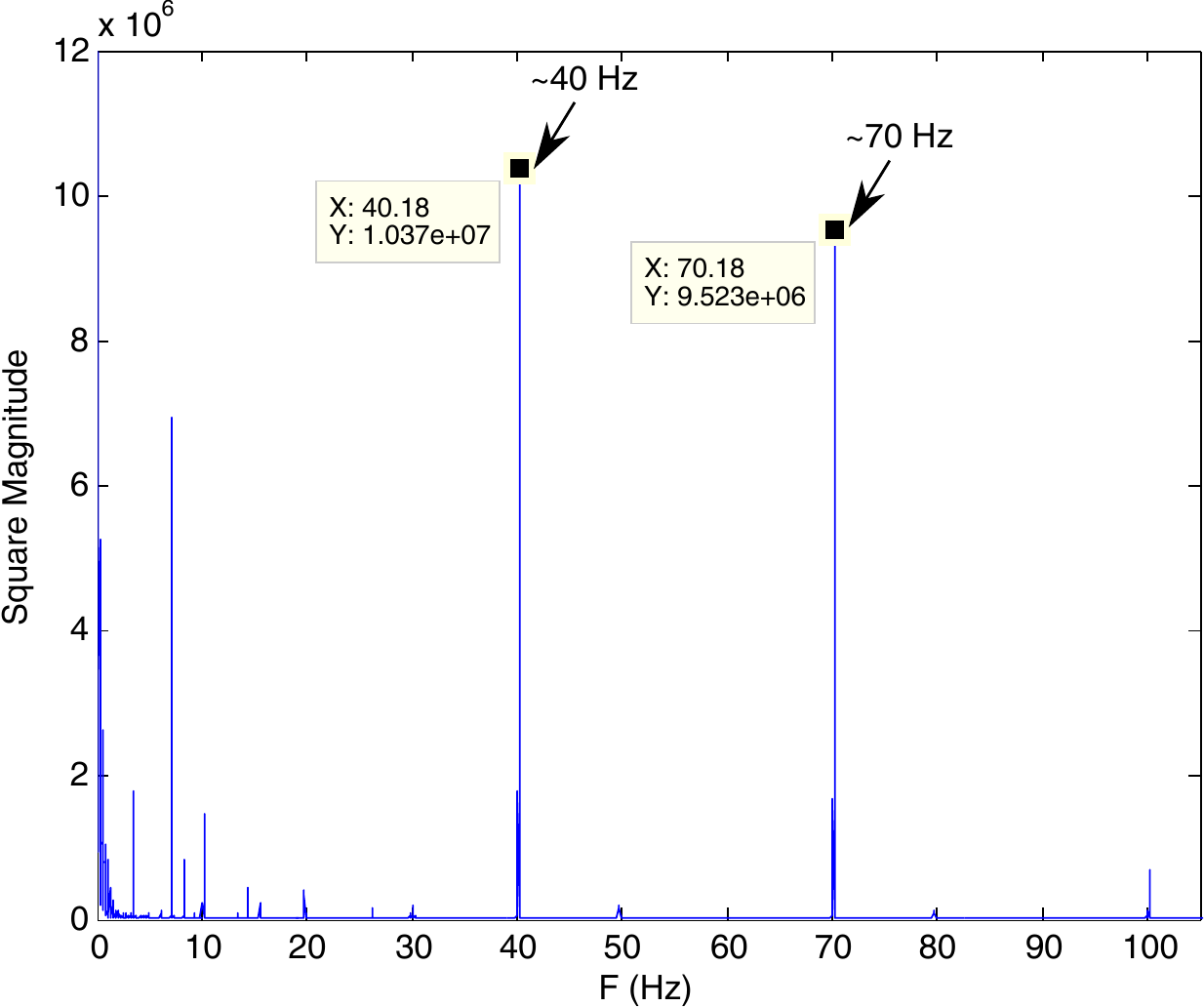}}
  %\hfill
  %\qquad
  \hskip 6em
  \subfigure[]{\includegraphics[width=67mm]{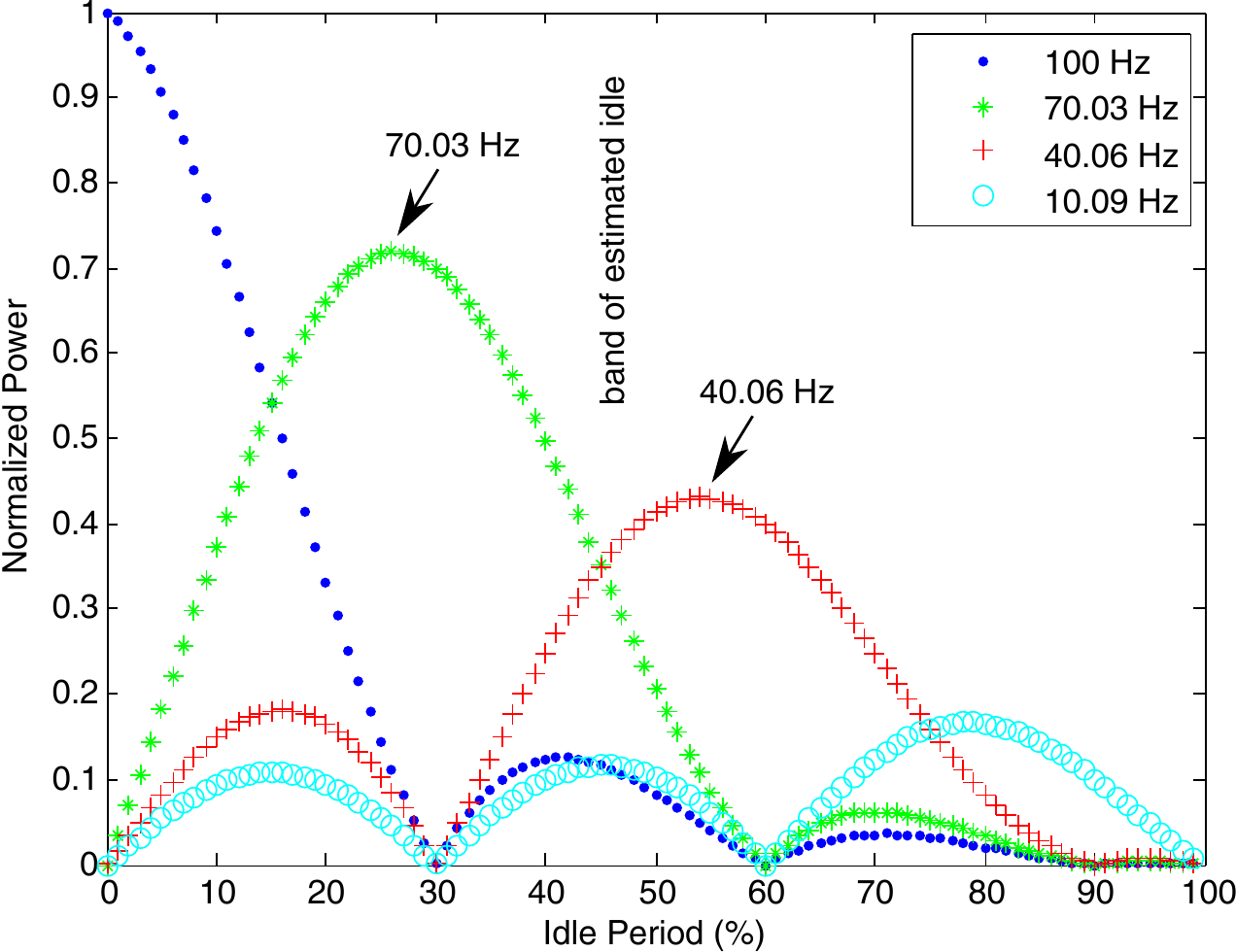}}
  \caption{(a) Frequency Spectrum for the video with 30 fps in 480P resolutions captured by CanonPowerShot SX230 HS, video-1. (b) The reference model - variation \textcolor{black}{in frequency of main ENF harmonic} vs. idle period for the same video. The estimated idle is in the range between 45\% and 50\% (The measured idle \cite{Hajj-Ahmad2016-ENF} is 48\%).}
  \label{Fig: Harmonic comparisons for video1}
\end{figure*}

\begin{figure*}[t]
  \centering
  \subfigure[]{\includegraphics[width=64mm]{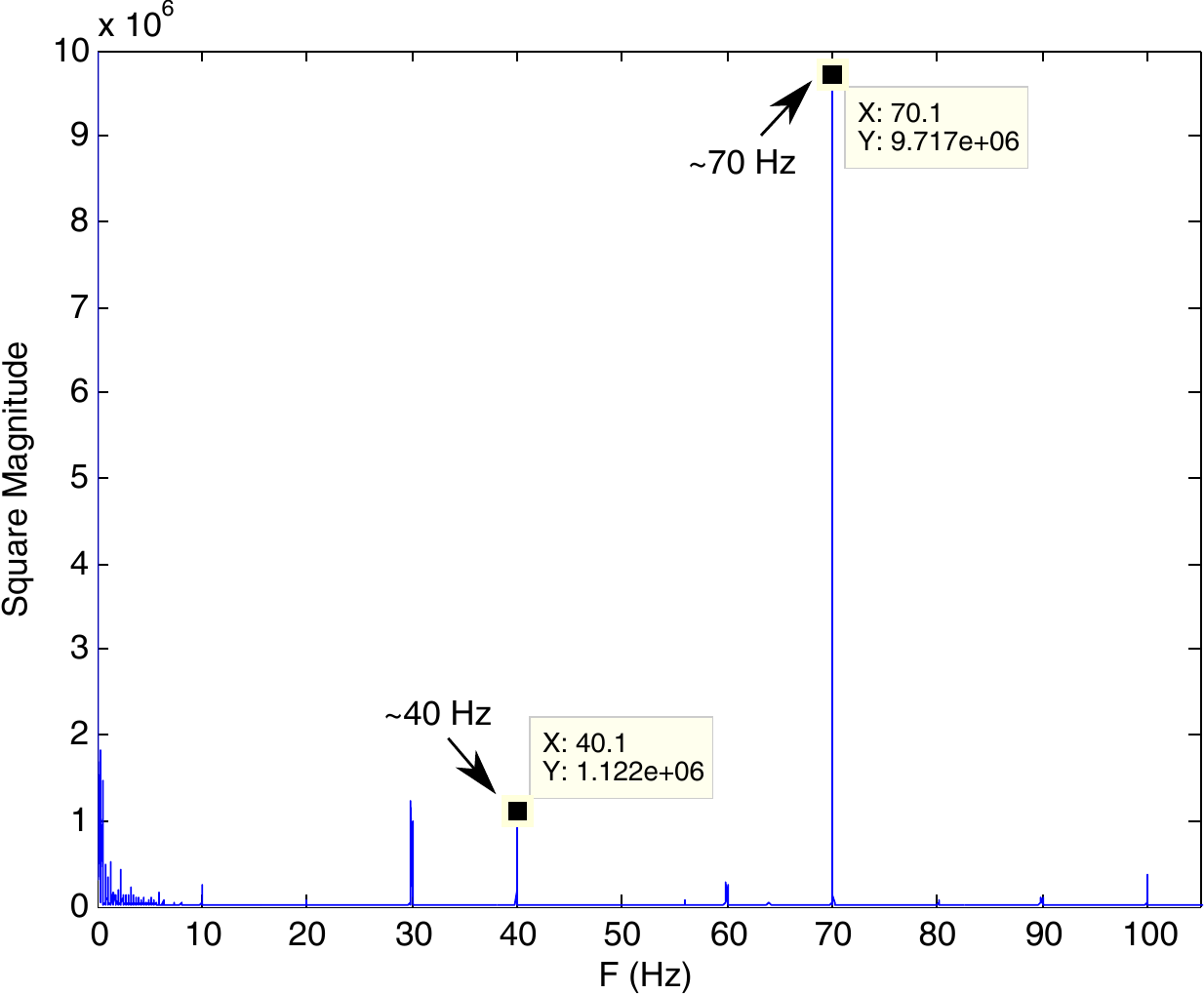}}
  %\hfill
  %\qquad
  \hskip 6em
  \subfigure[]{\includegraphics[width=66mm]{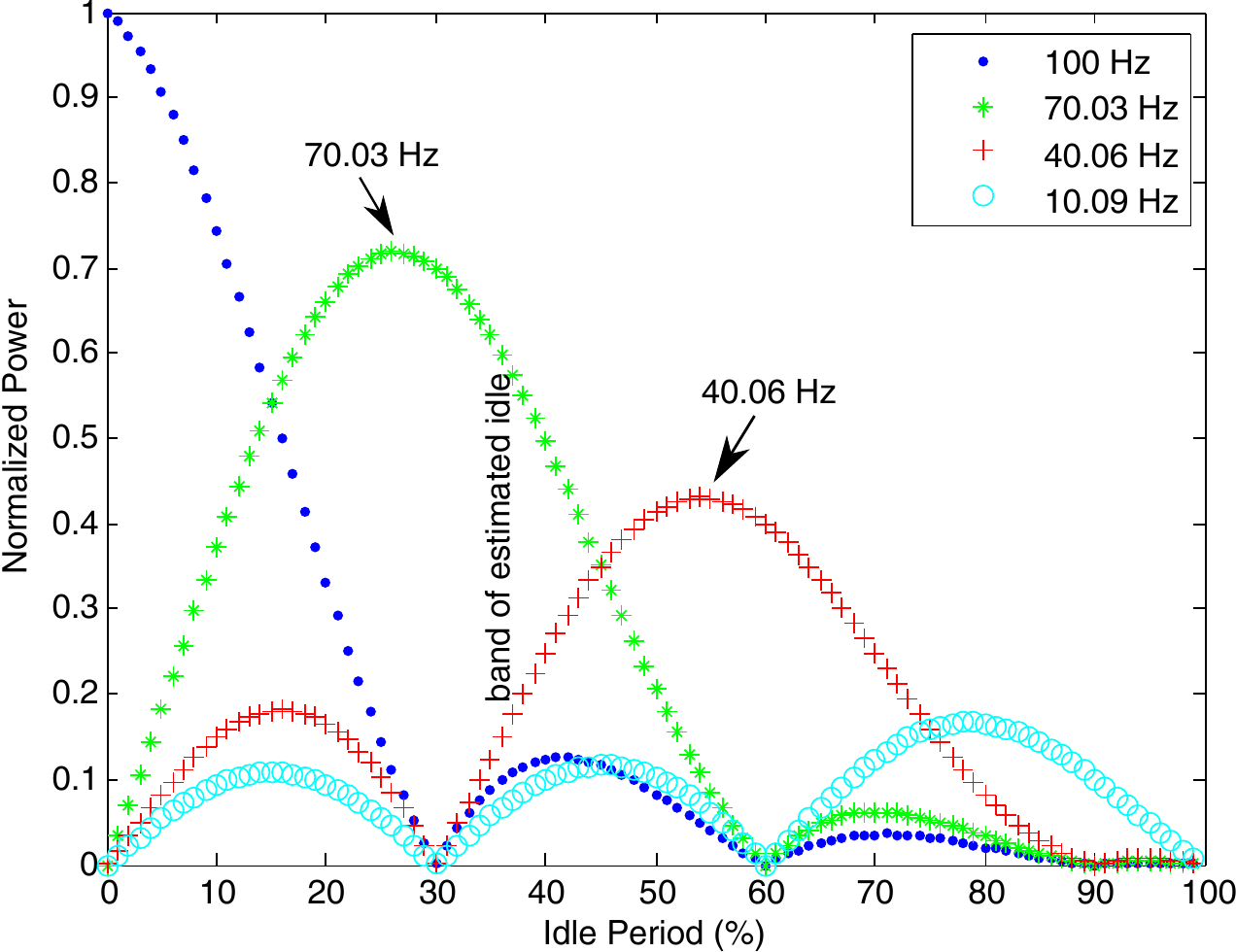}}
  \caption{(a) Frequency Spectrum for the video with 30 fps in 720P resolutions captured by CanonPowerShot SX230 HS, video-2. (b) The reference model - variation \textcolor{black}{in frequency of main ENF harmonic} vs. idle period for the same video. The estimated idle is in the range between 35\% and 40\% (The measured idle \cite{Hajj-Ahmad2016-ENF} is 38\%).}
  \label{Fig: Harmonic comparisons for video2}
\end{figure*}

\begin{figure*}[t]
  \centering
  \subfigure[]{\includegraphics[width=64mm]{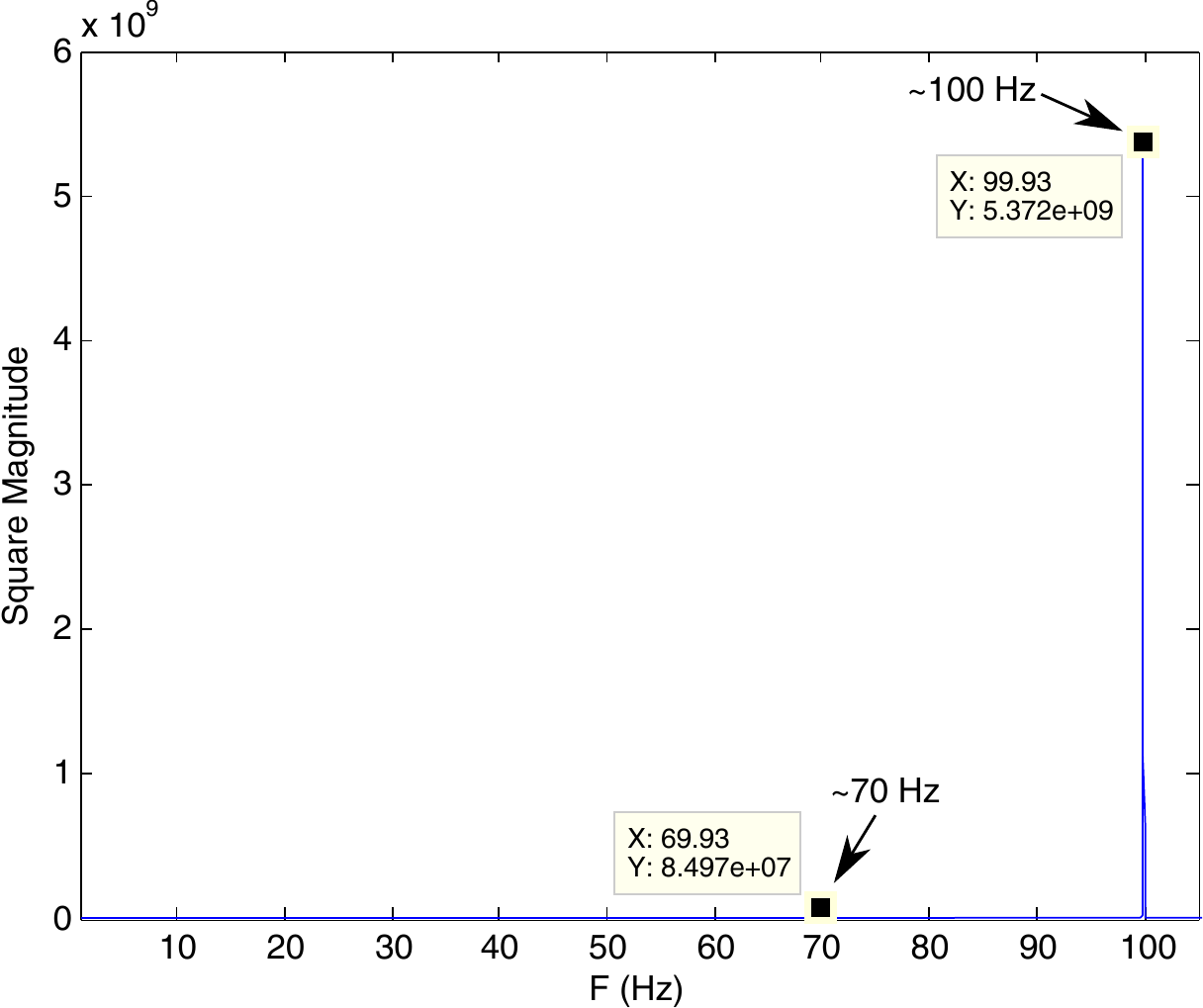}}
  %\hfill
  %\qquad
  \hskip 6em
  \subfigure[]{\includegraphics[width=67mm]{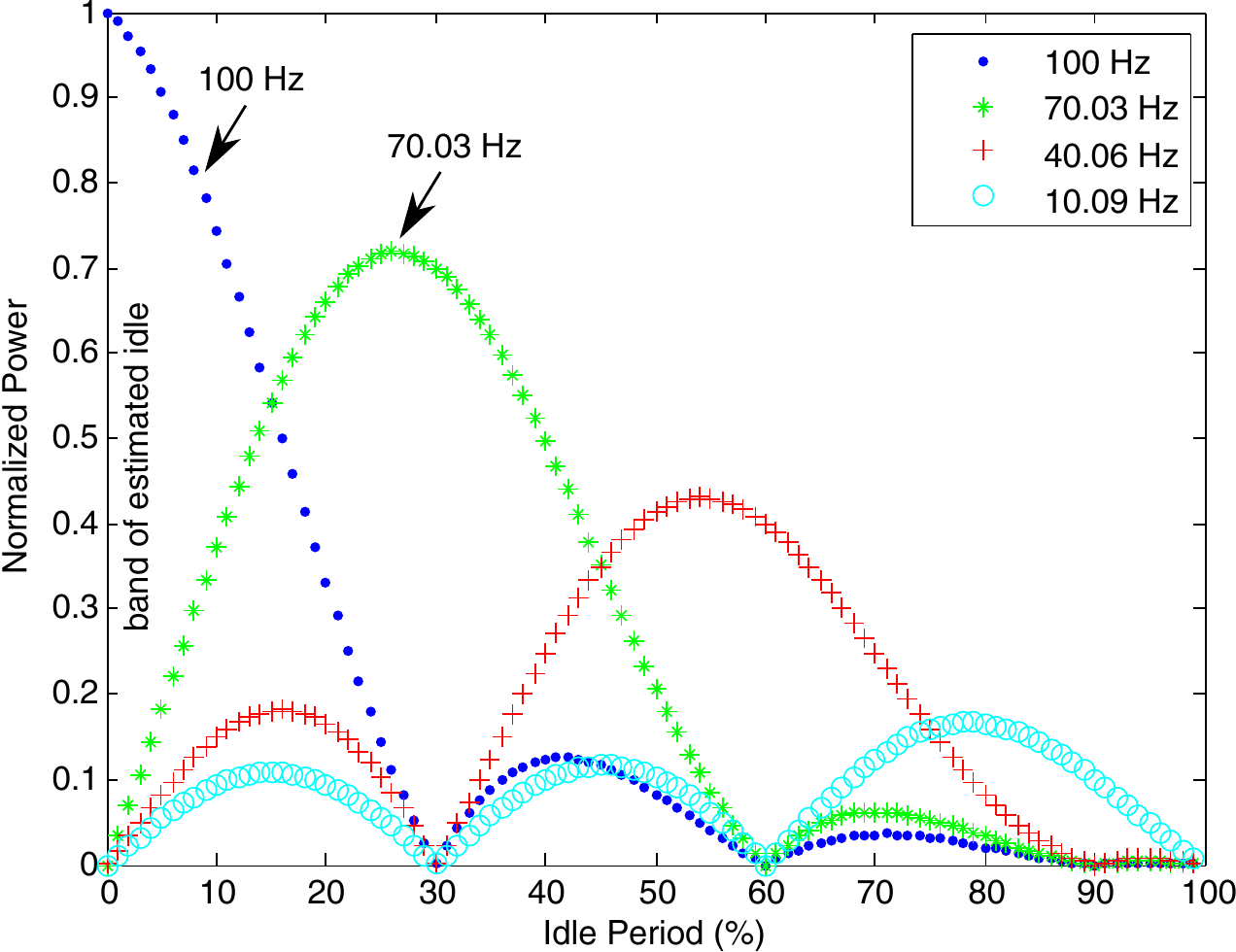}}
  \caption{(a) Frequency Spectrum for the video with 30 fps in 720P resolutions captured by Nikon D3100, video-3. (b) The reference model - variation \textcolor{black}{in frequency of main ENF harmonic} vs. idle period for the same video. The estimated idle is in the range between 0\% and 5\% (The measured idle \cite{Hajj-Ahmad2016-ENF} is 4\%).}
  \label{Fig: Harmonic comparisons for video3}
\end{figure*}

\begin{figure*}[t]
  \centering
  \subfigure[]{\includegraphics[width=64mm]{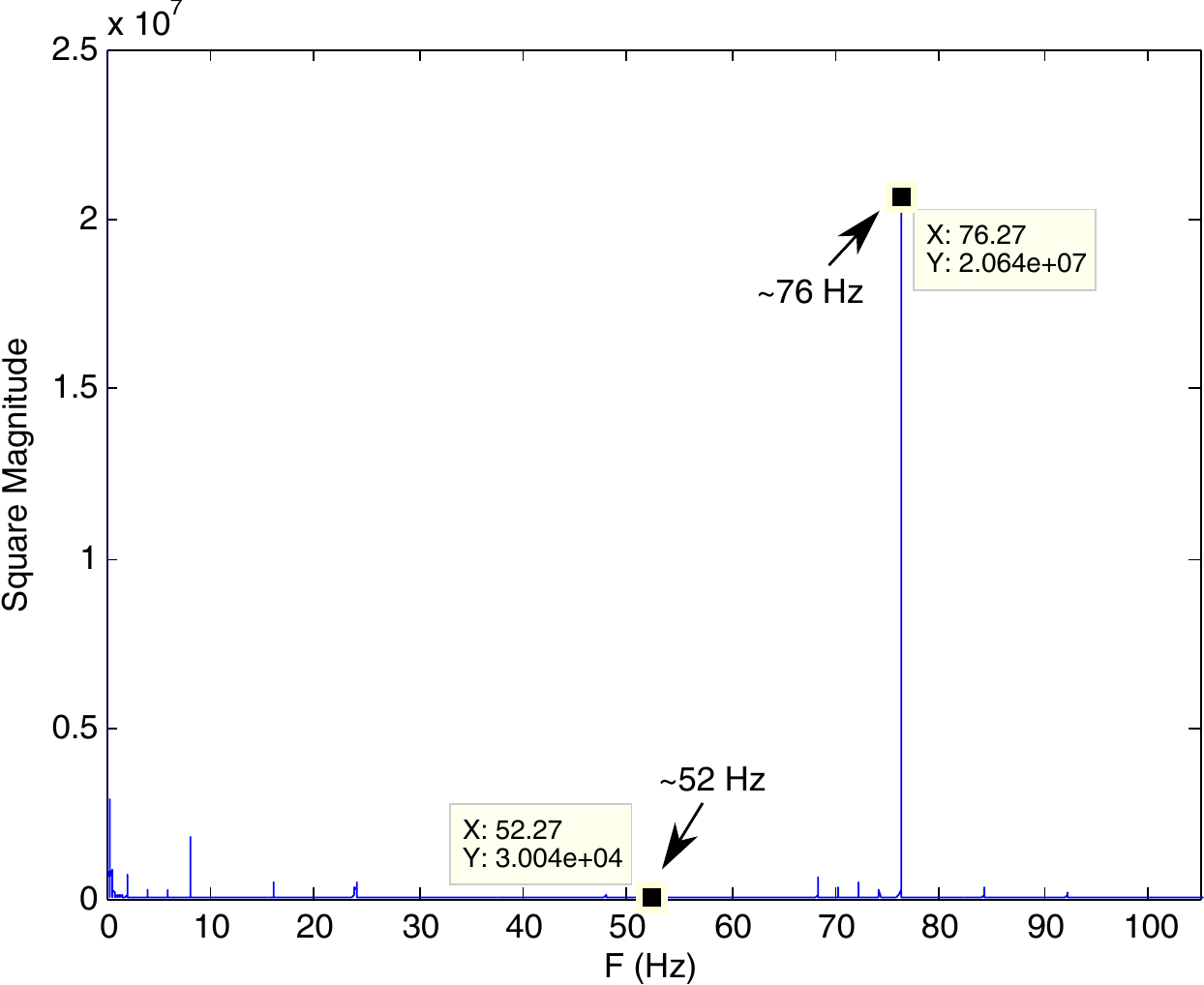}}
  %\hfill
  %\qquad
  \hskip 6em
  \subfigure[]{\includegraphics[width=65mm]{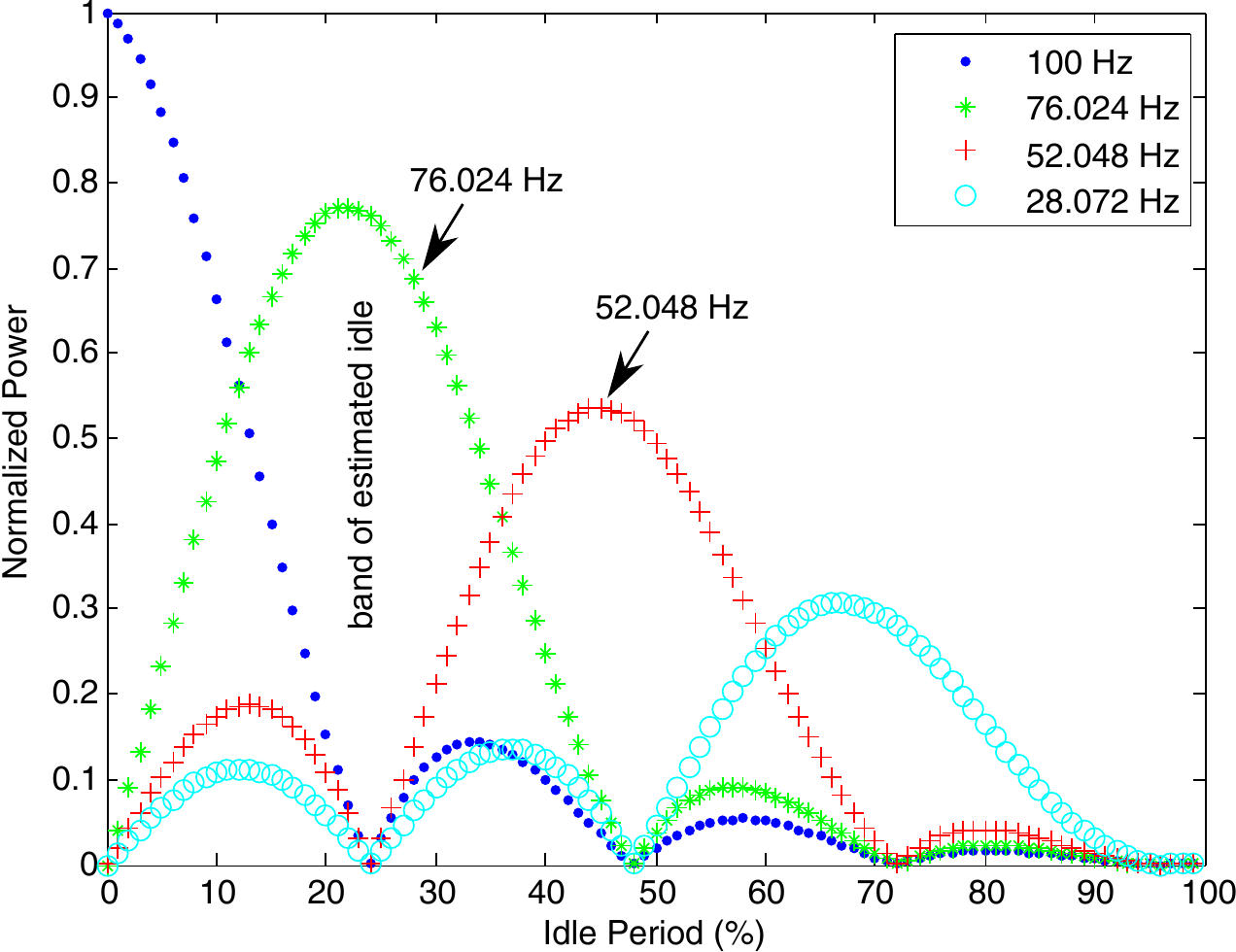}}
  \caption{(a) Frequency Spectrum for the video with 23.976 fps in 720P resolutions captured by Nikon D3100, video-4. (b) The reference model - variation \textcolor{black}{in frequency of main ENF harmonic} vs. idle period for the same video. The estimated idle is in the range between 20\% and 25\% (The measured idle \cite{Hajj-Ahmad2016-ENF} is 23\%).}
  \label{Fig: Harmonic comparisons for video4}
\end{figure*}

%\begin{figure}[!t]
%\centering
%{\includegraphics[width=67mm]{_DSC0340_FrequencySpectrum_v2}}
%\caption{(a) Frequency Spectrum for a video with 23.976 fps in 720P resolutions captured by Nikon D3100, video-5. (b) The reference model - variation of ENF Harmonic vs. duration of idle period for the same video, video-5. The estimated idle is in the range between 20\% and 25\% (The measured idle \cite{Hajj-Ahmad2016-ENF} is 23\%).}
%\label{Fig: Harmonic comparisons for video5}
%\end{figure}

\begin{figure*}[t]
  \centering
  \subfigure[]{\includegraphics[width=64mm]{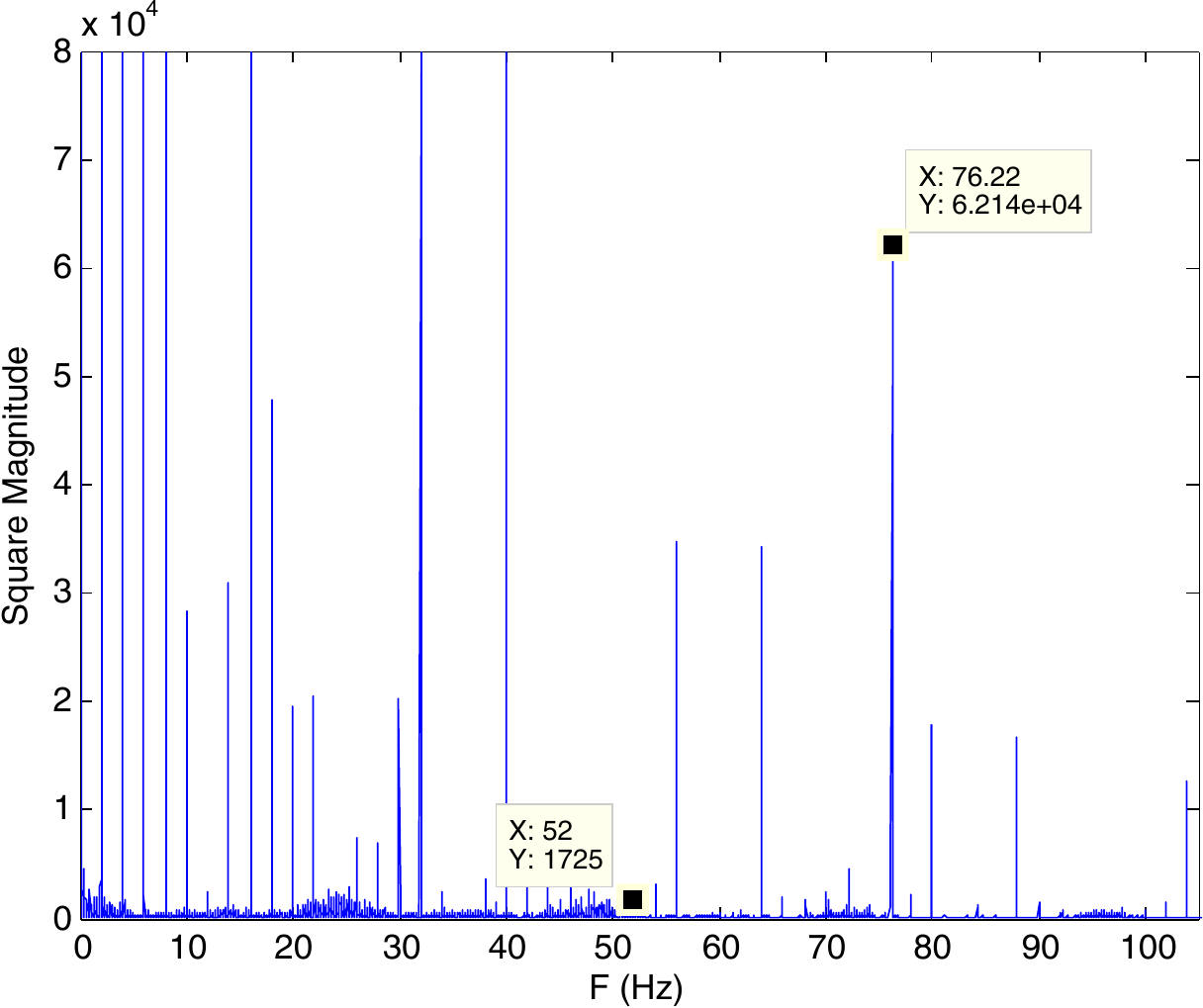}}
  %\hfill
  %\qquad
  \hskip 6em
  \subfigure[]{\includegraphics[width=67mm]{23_976fps_720P}}
  \caption{(a) Frequency Spectrum for the another 23.976 fps in 720P video by Nikon D3100, video-5 (captured under \textbf{CFL} bulb). (b) The reference model - variation \textcolor{black}{in frequency of main ENF harmonic} vs. idle period for the same video. The estimated idle is in the range between 20\% and 25\% (The measured idle \cite{Hajj-Ahmad2016-ENF} is 23\%).}
  \label{Fig: Harmonic comparisons for video5}
\end{figure*}

\begin{figure*}[t]
  \centering
  \subfigure[]{\includegraphics[width=66mm]{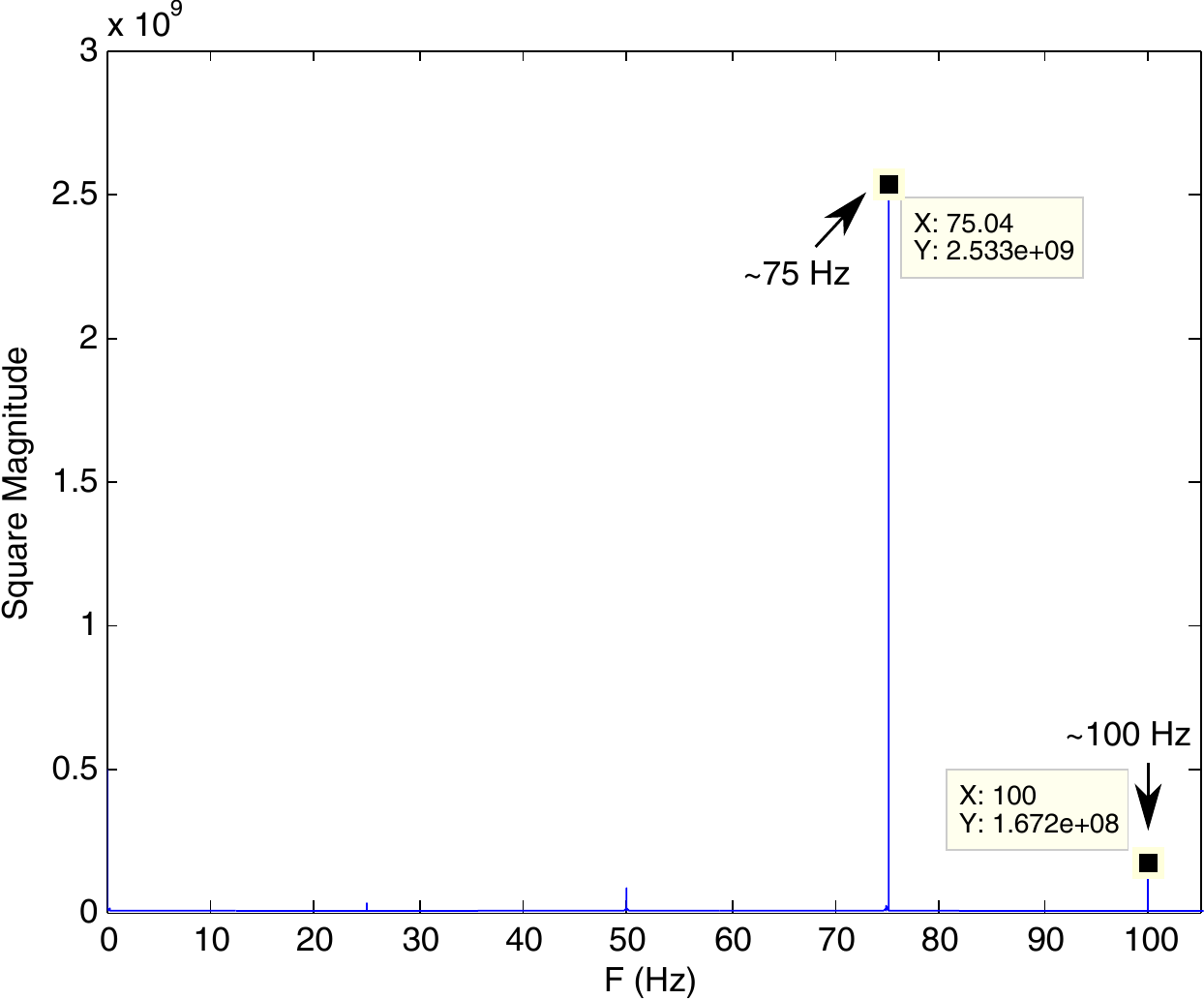}}
  %\hfill
  %\qquad
  \hskip 6em
  \subfigure[]{\includegraphics[width=67mm]{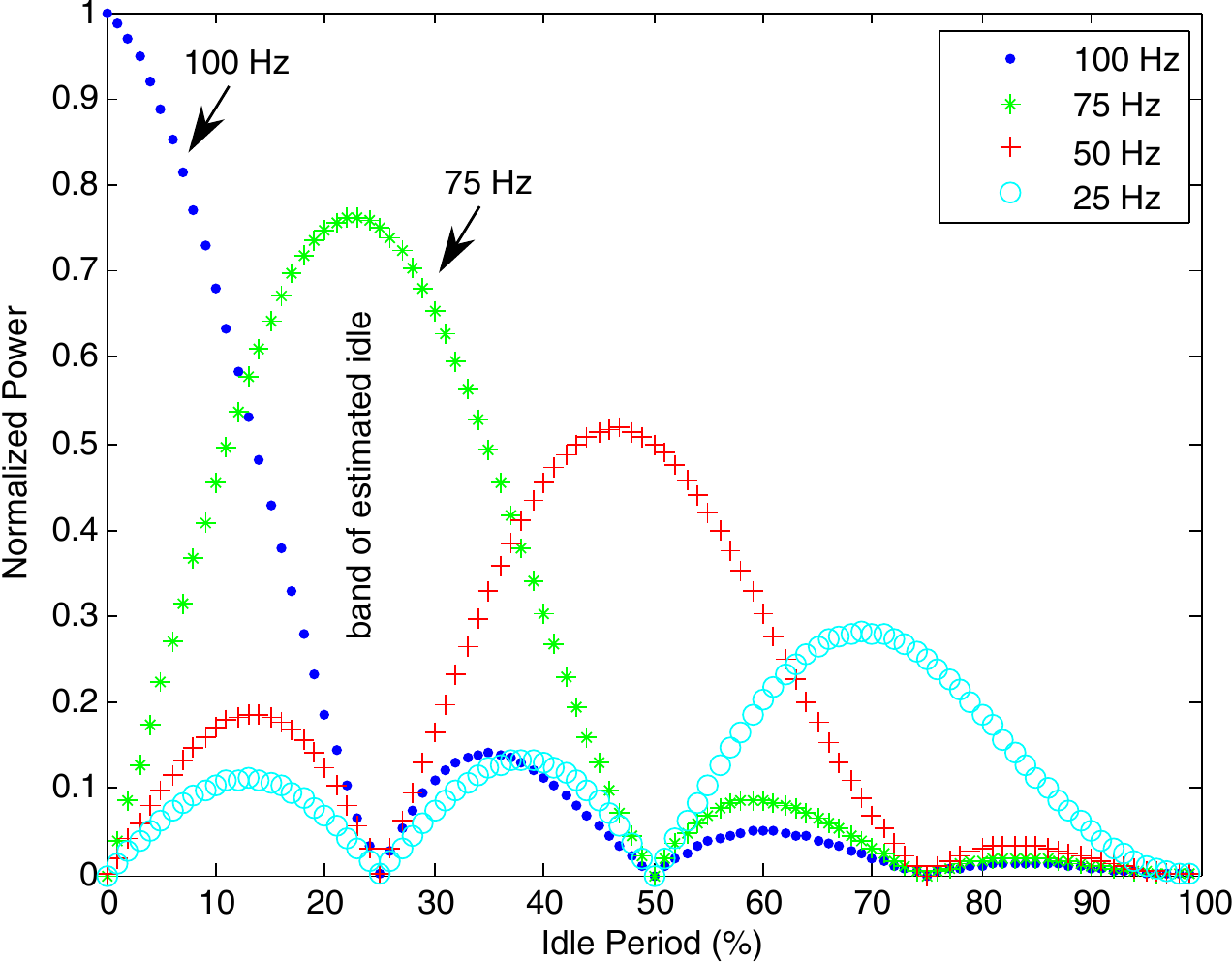}}
  \caption{(a) Frequency Spectrum for the video with 25 fps in 720P resolutions captured by Nikon D3100, video-6. (b) The reference model - variation \textcolor{black}{in frequency of main ENF harmonic} vs. idle period for the same video. The estimated idle period is in the range between 20\% and 25\%. Since the video frame rate is half the nominal ENF, i.e.\textcolor{black}{,} 0 Hz alias ENF, idle period cannot be computed via \cite{Hajj-Ahmad2016-ENF} for this case.}
  \label{Fig: Harmonic comparisons for video6}
\end{figure*}

\subsection{Proposed Approach}
\label{Section: Proposed Source Camera Verification}
In this subsection, a novel idle period \textcolor{black}{estimation} method for} videos containing an ENF signal, that can handle the  limitations of the state-of-the-art is proposed. The proposed method relies on the analytical model developed in section \ref{AnalyticalModel} and so is for videos exposed by a rolling shutter mechanism. According to the model, the nominal illumination harmonic, 100/120 Hz (\textcolor{black}{twice} the nominal ENF) is shifted to some other \textcolor{black}{frequency} depending on the idle period length as illustrated in Fig. \ref{idle_vs_ENF_for_30fp_50Hz_Model&Sim}. As the idle period increases, \textcolor{black}{captured power of the strongest ENF component} decreases, while the second strongest increases. For the proposed approach, the locations of these two \textcolor{black}{ENF components} as well as the power ratio between them form the key features. 

%Accordingly, the main steps of the proposed approach are  provided in Table \ref{Table_SourceCameraVerification}.
\textcolor{black}{The proposed approach consists of a number of operational steps}. First, the reference analytical model for the test video showing the variation \textcolor{black}{in frequency of the most powerful ENF component} depending on the possible idle period is derived based on the video frame rate and the nominal ENF in the region, where the video is recorded, as described in section \ref{AnalyticalModel}. Following this, the time-series for illumination variation throughout the video is formed by concatenating  the row illumination samples, one illumination sample per row via the steady pixels average, similar to the procedure in Fig. \ref{Fig_FilterBank} (a). For the details of how to form this time series, the reader is referred to \cite{su2014} and \cite{patent:20150356992}. Discrete Fourier Transform of this time-series is then computed. Based on the derived analytical model, the strongest two ENF \textcolor{black}{components} in the test video are located in the frequency (Fourier) domain. The ratio of the power of the located \textcolor{black}{ENF components} are also computed. Next, the corresponding idle point for these \textcolor{black}{ENF components} and their power ratio is located using the analytical model. Considering the noise factor, a small band in the vicinity of this point is assigned as candidate idle period range. The middle point of this range is taken as the estimated idle period.

Unlike to the state-of-the-art \cite{Hajj-Ahmad2016-ENF} which treats the sequence of mean luminance values of the $i^{th}$ rows of a sequence of frames of the video as a separate time series, the proposed approach concatenates the illumination samples of all rows of all consecutive frames. Hence, the sampling rate for the proposed approach is as high as \textit{video frame rate $\times$ number of rows} as discussed in section \ref{The Impact of Rolling Shutter}, which is much higher than the Nyquist Criteria, which leads the proposed approach to operate on videos with any frame rate, i.e.\textcolor{black}{,} no alias ENF.

The operational procedure of the proposed algorithm is clarified with the experiments in the next subsections. Besides,  results demonstrating how the proposed method is able to handle the limitations of the state-of-the art are provided.

% whereas it is as many as \textit{video frame rate} for the state-of-the-art. This fact

% in contrary to the state of the art, where the frame sampling rate is 

% Since there may be videos attributed to different cameras but have the same idle period, the proposed approach is more appropriate for verification rather than identification.

%These facts form the basis of the proposed approach.

%For a real video \textcolor{black}{(AHMET what is a real video??)}, also the same harmonics in the model can be concentrated. That is, for instance, for a video at 30 fps recorded in EU, only 100 Hz, 70 Hz, 40 Hz and 10 Hz harmonics can be analyzed.

% The point where a close relationship, in terms of location and the power ratio, is observed between the two greatest harmonics of the model and that of the real video may provide the idle period of the video.

\subsection{Experiments with Videos with Still Content}
\label{Source Camera Verification - still content videos}
In this subsection, the proposed \textcolor{black}{idle period estimation} approach is tested on wall-scene videos that are used in section \ref{State of the Art Approach}. The results are compared with that of the vertical phase analysis technique. Besides, a video with 25 fps (a divisor of the nominal ENF, 50 Hz) is tested, where the vertical phase analysis approach is unable to work.

A comparison between the frequency spectrum of a 480 P wall-scene video at 30 fps that was recorded by a Canon PowerShot SX230HS (video-1), and the corresponding model is shown in Fig. \ref{Fig: Harmonic comparisons for video1} (a) and (b), respectively. According to Fig. \ref{Fig: Harmonic comparisons for video1} (a), 40 Hz \textcolor{black}{ENF component} has the  highest  power and 70 Hz \textcolor{black}{ENF component} is the second highest. The power ratio between them is around 1.1. The corresponding \textcolor{black}{ENF components} in the model, Fig. \ref{Fig: Harmonic comparisons for video1} (b) are located between idle periods of 45\% and  50\%.  Referring back to Fig. \ref{Fig: vertical phase-1-2} (a), the idle period of the same video is estimated as 48\% based on  vertical phase analysis.

Similarly, a comparison between the frequency spectrum of a 720 P wall-scene video at 30 fps that was recorded by a Canon PowerShot SX230HS (video-2), and the corresponding model are shown in Fig. \ref{Fig: Harmonic comparisons for video2} (a) and (b), respectively. In Fig. \ref{Fig: Harmonic comparisons for video2} (a), 70 Hz \textcolor{black}{ENF component} is the highest and 40 Hz \textcolor{black}{ENF component} is the second highest. The power ratio between them is around 8.5. The corresponding \textcolor{black}{ENF components} in the model, Fig. \ref{Fig: Harmonic comparisons for video2} (b) are located between  idle periods of  35\% and 40\%. Referring back to Fig. \ref{Fig: vertical phase-1-2} (b), the idle period of the same video was estimated as 38\% based on vertical phase analysis. It is notable that as the proposed approach is independent of video resolution, as the reference model illustrations for video-1 and video-2 are the same as shown in Fig. \ref{Fig: Harmonic comparisons for video1} (b) and Fig. \ref{Fig: Harmonic comparisons for video2} (b).

Similar experiments and comparisons were conducted for a 720P video at 30 fps (video-3), and for a 720P video at 23.976 fps (video-4),  recorded by a Nikon D3100. The estimated idle period for video-3 is between 0\% and 5\%, whereas it is between 20\% and 25\% for video-4 as shown in Fig. \ref{Fig: Harmonic comparisons for video3} and Fig. \ref{Fig: Harmonic comparisons for video4}, respectively. Referring back to Fig. \ref{Fig: vertical phase-3-4} (a) and Fig. \ref{Fig: vertical phase-3-4} (b), the idle period of the same videos were estimated as 4\% and 23\% respectively for video-3 and video-4 based on the vertical phase method. It is notable that as the proposed approach is dependent on the video frame rate, reference model illustrations for video-3 and video-4 are not the same as shown in Fig. \ref{Fig: Harmonic comparisons for video3} (b) and Fig. \ref{Fig: Harmonic comparisons for video4} (b).

In Fig. \ref{Fig: Harmonic comparisons for video5} (a), the frequency spectrum of another 23.976 fps wall-scene video, 720 P, recorded by a Nikon D3100 is provided. This  video was recorded under CFL bulb illumination, in which the vertical phase analysis failed to work as discussed in section \ref{State of the Art Approach},  Fig. \ref{Fig: vertical phase-5 and frequency domain}. When compared to the corresponding model illustration in Fig. \ref{Fig: Harmonic comparisons for video5} (b), the idle period for this video is estimated between 20\% and 25\%, which is the same as that for video 4 illustrated in Fig. \ref{Fig: Harmonic comparisons for video4}. This outcome indicates that the proposed approach is more robust than the state-of-the-art for a noisy video where the power of the ENF signal is weak. This may be due to the reason that the power ratio between the strongest two ENF \textcolor{black}{components} is likely to be preserved, even though the power of the each \textcolor{black}{component} is reduced.

In Fig. \ref{Fig: Harmonic comparisons for video6} (a) and (b), respectively a comparison between frequency spectrum of 25 fps video in 720 P (video-6), recorded by a Nikon D3100 in a region where nominal ENF is 50 Hz and the corresponding model are provided. Based on the argument in this section, the idle period is estimated between 20\% and 25\%, which is a similar result to the case for 23.976 fps video, as shown in  Fig. \ref{Fig: Harmonic comparisons for video4} (a). However this time, when compared to Fig. \ref{Fig: Harmonic comparisons for video4} (a), the proportion of the power of the \textcolor{black}{strongest ENF component} to the second \textcolor{black}{strongest} is much lower, which corresponds to a smaller idle period. It is reasonable that idle period should be inversely proportional to video frame rate since the total time per frame should be smaller for videos with greater frame rate. Hence, it is an indication that the idle period for video-6 is closer to 20\% than that for video-4. This experiment shows that the proposed method works also for videos with a frame rate that is a divisor of the nominal ENF. Although highlighted in the previous sections that vertical phase analysis cannot work under such a frame rate, i.e.\textcolor{black}{,} a divisor of nominal ENF, it may be argued that testing a video with a frame rate in close proximity may provide an approximation to the expected idle period. However, most consumer cameras provide  very few options for frame rates. It should \textcolor{black}{also} be noted that neither the frame read-out time nor the idle period is  provided by camera manufacturers. This is why all the comparisons in this subsection and also in the next are made \textcolor{black}{by using} vertical phase analysis.

\begin{table}[t]
%\begin{table}[ht]
\caption{Tabulated findings of the proposed model for variation \textcolor{black}{in frequency of main ENF harmonic} vs. idle period for a 29.97 fps video recorded in EU\textcolor{black}{.}}
\centering
  \begin{tabular}{c c c c c c}
  	\hline\hline
	Idle Period (\%) & $H_1$ (Hz) &  $H_2$ (Hz) & $P_{H_1}/P_{H_2}$ \bigstrut\\
	\hline
	0 & 100 & 0 & $\approx \infty$ \bigstrut\\
	
	5 & 100 & 70 & $5.0$ \bigstrut\\
	
	10 & 100 & 70 & $2.0$ \bigstrut\\
	
	15 & 100, 70 & 100, 70 & $1.0$ \bigstrut\\
	
	20 & 70 & 100 & $2.0$ \bigstrut\\
	
	25 & 70 & 100 & $5.0$ \bigstrut\\
	
	30 & 70 & 100, 40 & $505.6$ \bigstrut\\
	
	35 & 70 & 40 & $5.0$ \bigstrut\\
	
	40 & 70 & 40 & $2.0$ \bigstrut\\
	
	45 & 70, 40 & 70, 40 & $1.0$ \bigstrut\\
	
	50 & 40 & 70 & $2.0$ \bigstrut\\
	
	55 & 40 & 70 & $5.0$ \bigstrut\\
	
	60 & 40 & 10, 70 & $498.5$ \bigstrut\\
	
	65 & 40 & 10 & $4.9$ \bigstrut\\
	
	70 & 40 & 10 & $2.0$ \bigstrut\\
	
	75 & \textcolor{black}{40, 10} & \textcolor{black}{40, 10} & $1.0$ \bigstrut\\
	
	80 & 10 & 40 & $2.0$ \bigstrut\\
	
	85 & 10 & 40 & $5.1$ \bigstrut\\
	
	90 & 10 & 40 & $334.0$ \bigstrut\\
	
	95 & 10 & 40 & $6.9$ \bigstrut\\
	
    \hline\hline
  \end{tabular}
\label{Table: Harmonic vs idle for 30 fps}
\end{table}

\begin{table}[t]
%\begin{table}[ht]
\caption{Tabulated findings of the proposed model for variation \textcolor{black}{in frequency of main ENF harmonic} vs. idle period for a 25 fps video recorded in EU\textcolor{black}{.}}
\centering
  \begin{tabular}{c c c c c c}
  	\hline\hline
	Idle Period (\%) & $H_1$ (Hz) &  $H_2$ (Hz) & $P_{H_1}/P_{H_2}$ \bigstrut\\
	\hline
	0 & 100 & 0 & $\approx \infty$ \bigstrut\\
	
	5 & 100 & 75 & $4.0$ \bigstrut\\
	
	10 & 100 & 75 & $1.5$ \bigstrut\\
	
	15 & 75 & 100 & $1.5$ \bigstrut\\
	
	20 & 75 & 100 & $4.0$ \bigstrut\\
	
	25 & 75 & 100 & $\approx \infty$ \bigstrut\\
	
	30 & 75 & 50 & $4.0$ \bigstrut\\
	
	35 & 75 & 50 & $1.5$ \bigstrut\\
	
	40 & 50 & 75 & $1.5$ \bigstrut\\
	
	45 & 50 & 75 & $4.0$ \bigstrut\\
	
	50 & 50 & 100 & $\approx \infty$ \bigstrut\\
	
	55 & 50 & 25 & $4.0$ \bigstrut\\
	
	60 & 50 & 25 & $1.5$ \bigstrut\\
	
	65 & 25 & 50 & $1.5$ \bigstrut\\
	
	70 & 25 & 50 & $4.0$ \bigstrut\\
	
	75 & 25 & 75 & $\approx \infty$ \bigstrut\\
	
	80 & 25 & 50 & $6.0$ \bigstrut\\
	
	85 & 25 & 50 & $3.5$ \bigstrut\\
	
	90 & 25 & 50 & $2.7$ \bigstrut\\
	
	95 & 25 & 50 & $2.2$ \bigstrut\\
	
    \hline\hline
  \end{tabular}
\label{Table: Harmonic vs idle for 25 fps}
\end{table}

\begin{figure}[t]
\centering
\includegraphics[width=76mm]{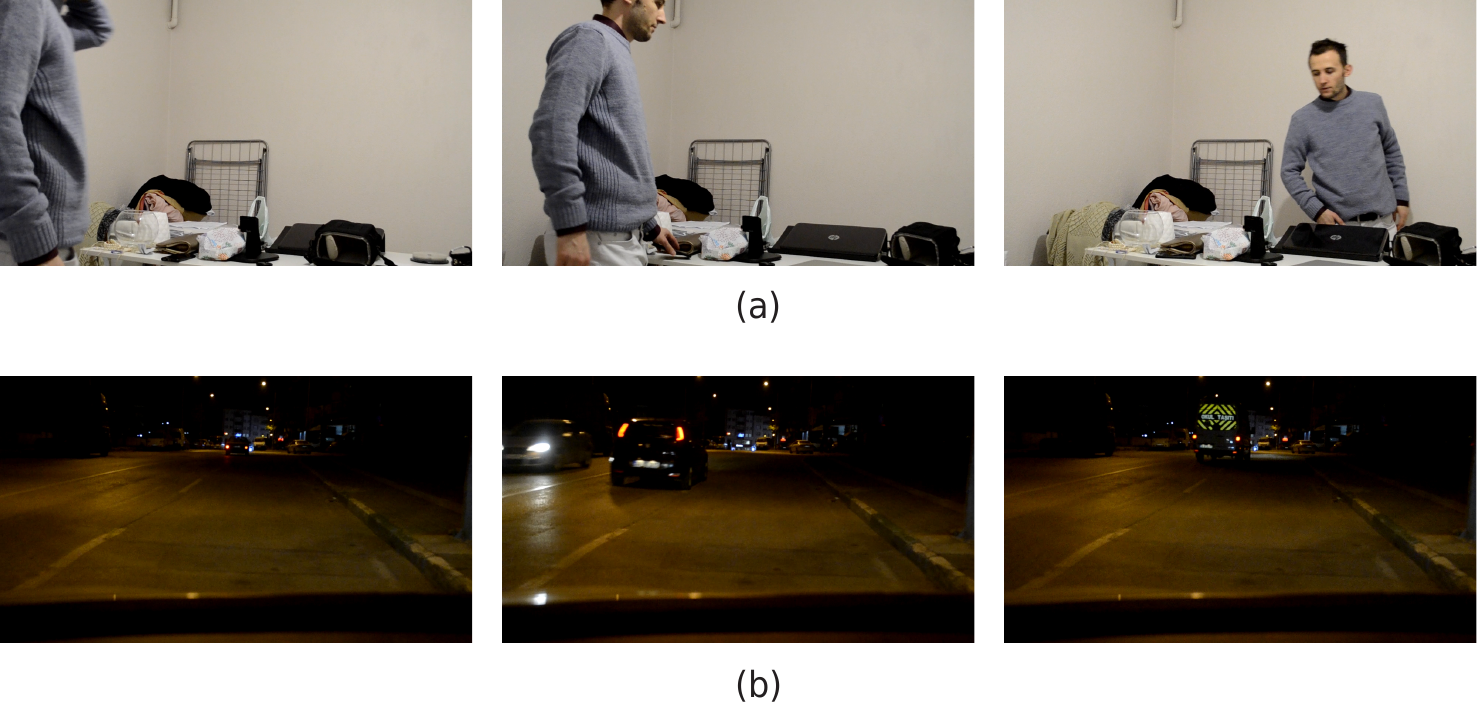}
%\vspace{2mm}
\caption{\textcolor{black}{Sample frames from videos with moving content (a) an indoor video (b) an outdoor video}\textcolor{black}{.}}
\label{SampleFrames}
%\vspace{2mm}
\end{figure}

\subsection{Experiments with Videos with Moving Content}
\label{section:Experiment-source camera verification with moving content}

In this subsection, the performance of \textcolor{black}{the proposed idle period estimation technique is evaluated by conducting experiments on videos with moving content that are captured using 5 different fixed cameras. Each dataset for each camera included 6 different videos, half of which were indoor and the other half outdoor. All the videos were captured in Turkey, where the nominal ENF is 50 Hz. \textcolor{black}{Since the power of the estimated ENF signal in a video, and so the performance of the proposed approach may be affected from the type of mains-powered light source, and the power of illumination emitted by mains-powered light source in relative to that of the non-mains-powered ones in the scene, various settings were used to build the dataset. Both the indoor and the outdoor videos were recorded under different light sources such as LED bulb, CFL bulb and sodium vapor light. Most of the outdoor videos were captured in poorly illuminated streets.} Sample frames from 2 exemplary videos for indoor and outdoor scenes respectively are provided in Fig. \ref{SampleFrames} (a) and (b)}.

\textcolor{black}{First, the results for each of the videos from two different cameras were analyzed providing the estimated key parameters. The first dataset included 480 P videos at 29.97 fps recorded by a Canon PowerShot SX230HS model camera, and the second dataset had 25 fps video recorded at 720 P by a Nikon D3100 model camera.}
%The first dataset included 720 P videos at 29.97 fps recorded by a Nikon D3100 model camera. The second dataset has 25 fps video recorded at 720 P by the same Nikon camera. The third dataset has 720 P videos recorded at 29.97 fps by a Canon PowerShot SX230HS camera and the fourth dataset is composed of 480 P videos at  29.97 fps captured by the same Canon camera.
%From the previous analysis, it should be recalled that idle period for a camera is likely to vary depending on video frame rate and resolution. Hence, the 4 video datasets should have different idle periods. Nevertheless, as the proposed model for the variation of ENF harmonic depending on the idle period length is independent of video resolution, 2 different reference model illustrations only are obtained and used for comparisons with these 4 different cases.
Table \ref{Table: Harmonic vs idle for 30 fps} and Table \ref{Table: Harmonic vs idle for 25 fps} show the largest 2 \textcolor{black}{ENF components} and the power ratio between them, at idle periods of multiples of 5\%, for 29.97 fps and for 25 fps videos respectively, which are extracted from the analytical model illustrations in Fig. \ref{Fig: Harmonic comparisons for video3} and in Fig. \ref{Fig: Harmonic comparisons for video6}, respectively. They form the reference parameters for comparisons with those of the test videos. {Table \ref{Table: Harmonic vs idle for tested videos} provides the main metadata for each test video as well as the location of the largest two \textcolor{black}{ENF components} and the power ratio between them  obtained via Fourier analysis and from the proposed analytical model. Table \ref{Table: Harmonic vs idle for tested videos} also provides the estimated idle period, the expected idle period and the relative error for each video. Estimated idle period is obtained based on the search of the corresponding match between the reference parameters and the test video parameters. The expected idle period for \textcolor{black}{the 30 fps videos} was obtained from still-content videos in section \ref{State of the Art Approach} by using vertical phase analysis \cite{Hajj-Ahmad2016-ENF}. The expected idle period for the 25 fps video set is obtained in section \ref{Source Camera Verification - still content videos} \textcolor{black}{mainly} based on the proposed approach. Since a 25 fps video set is a divisor of nominal ENF, the vertical phase analysis approach is unable to work for this condition. It should be emphasized also that idle period information is \textcolor{black}{unlikely to be} provided by the camera manufacturers}. According to the Table \ref{Table: Harmonic vs idle for tested videos}, the idle period of \textcolor{black}{10 videos out of 12, i.e.\textcolor{black}{,} 83\%, were successfully estimated in the vicinity of $\pm$ 5\% of the expected idle period. The idle period of 1 video was estimated around the vicinity of $\pm$ 10\% of the expected idle period, though the test fails for 1 video, where no match was found.}

\textcolor{black}{Second, idle period estimation statistics for the videos in the same settings, yet captured by 5 different cameras, 6 videos for each camera, were explored. The cameras were GoPro Hero 4, Nikon D3100, Nikon P100, Canon SX230HS and Canon SX220HS. Each video by each camera model was captured at 720 P and at 29.97 fps. Table \ref{Table:Statistics_for_idle} provides the median value of estimated idle periods for the videos by each camera and average estimation error based on the estimated median. Accordingly, the median value for the each camera is very close to the expected idle period, and the average estimation error is in close vicinity of the median value. From the table, it can also be noticed that the idle period for Nikon P100, Canon SX230HS and Canon SX220HS cameras are very close to each other. Hence,  different cameras may have similar or the same idle period.}

\textcolor{black}{One possible application of the idle period in camera forensics may be verifying if 2 videos were produced by different cameras. That is, if the idle periods of the two videos are very similar, it may not be an indication that they are captured with the same camera. However, if the idle periods of the videos are distinct from each other, it is most likely that the videos were captured by different cameras.}

\begin{table*}[t]
%\begin{table}[ht]
\caption{The settings and estimated outcomes of each video used for evaluation of the proposed \textcolor{black}{idle period estimation} method\textcolor{black}{.}}
\centering
  \begin{tabular}{c c c c c c c c c c c c}
  	\hline\hline
	Camera Model & Vid. no & Res. (P) & Fr. rate (fps) & $H_1$ (Hz) &  $H_2$ (Hz) & $P_{H_1}/P_{H_2}$ & Estimated idle (\%) & \textcolor{black}{Expected} idle (\%) & Err (\%) \bigstrut\\
	\hline
%	Nikon D3100 & 1 & 720 & 29.97 & 100 & 70 & $7.7$ & $\textcolor{black}{2.5}$ & 4 & \textbf{1.5}  \bigstrut\\
%	
%	Nikon D3100 & 2 & 720 & 29.97 & 100 & 70 & $1.4$ & $\textcolor{black}{12.5}$ & 4 & 8.5 \bigstrut\\
%	
%	Nikon D3100 & 3 & 720 & 29.97 & 100 & 70 & $2.2$ & $\textcolor{black}{7.5}$ & 4 & \textbf{3.5} \bigstrut\\
%	
%	Nikon D3100 & 4 & 720 & 29.97 & 100 & 70 & $1.3$ & $\textcolor{black}{12.5}$ & 4 & 8.5 \bigstrut\\
%	
%	Nikon D3100 & 5 & 720 & 29.97 & 100 & 70 & $39.8$ & $\textcolor{black}{2.5}$ & 4 & \textbf{1.5} \bigstrut\\
%	
%	Nikon D3100 & 6 & 720 & 29.97 & 100 & 70 & $37.7$ & $\textcolor{black}{2.5}$ & 4 & \textbf{1.5} \bigstrut\\
%	
%	\hline
%	
%	
%	Canon SX230HS & 13 & 720 & 29.97 & 70 & 100 & $20.9$ & $\textcolor{black}{27.5}$ & 38.0 & 10.5 \bigstrut\\
%	
%	Canon SX230HS & 14 & 720 & 29.97 & 70 & 40 & $3.8$ & $\textcolor{black}{37.5}$ & 38.0 & \textbf{0.5} \bigstrut\\
%	
%	Canon SX230HS & 15 & 720 & 29.97 & 70 & 40 & $1.5$ & $\textcolor{black}{42.5}$ & 38.0 & \textbf{4.5} \bigstrut\\
%	
%	Canon SX230HS & 16 & 720 & 29.97 & 70 & 40 & $2.2$ & $\textcolor{black}{37.5}$ & 38.0 & \textbf{0.5} \bigstrut\\
%	
%	Canon SX230HS & 17 & 720 & 29.97 & 70 & 40 & $4.3$ & $\textcolor{black}{37.5}$ & 38.0 & \textbf{0.5} \bigstrut\\
%	
%	Canon SX230HS & 18 & 720 & 29.97 & 70 & 40 & $3.4$ & $\textcolor{black}{37.5}$ & 38.0 & \textbf{0.5} \bigstrut\\
%	
%	\hline
	
	Canon SX230HS & 1 & 480 & 29.97 & 40 & 70 & $1.9$ & $\textcolor{black}{47.5}$ & 48.0 & \textbf{0.5} \bigstrut\\
	
	Canon SX230HS & 2 & 480 & 29.97 & 70 & 40 & $1.0$ & $\textcolor{black}{45}$ & 48.0 & \textbf{3.0} \bigstrut\\
	
	Canon SX230HS & 3 & 480 & 29.97 & 70 & 40 & $1.1$ & $\textcolor{black}{47.5}$ & 48.0 & \textbf{0.5} \bigstrut\\
	
	Canon SX230HS & 4 & 480 & 29.97 & 40 & 70 & $1.27$ & $\textcolor{black}{47.5}$ & 48.0 & \textbf{0.5} \bigstrut\\
	
	Canon SX230HS & 5 & 480 & 29.97 & 40 & 70 & $1.21$ & $\textcolor{black}{47.5}$ & 48.0 & \textbf{0.5} \bigstrut\\
	
	Canon SX230HS & 6 & 480 & 29.97 & 40 & 70 & $1.4$ & $\textcolor{black}{47.5}$ & 48.0 & \textbf{0.5} \bigstrut\\
	
	\hline
	
	Nikon D3100 & 7 & 720 & 25 & 75 & 100 & $3.5$ & $\textcolor{black}{17.5}$ & \textcolor{black}{21.5} & \textbf{4.0} \bigstrut\\
	
	Nikon D3100 & 8 & 720 & 25 & 75 & 100 & $3.6$ & $\textcolor{black}{17.5}$ & \textcolor{black}{21.5} & \textbf{4.0} \bigstrut\\
	
	Nikon D3100 & 9 & 720 & 25 & 75 & 100 & $2.1$ & $\textcolor{black}{17.5}$ & \textcolor{black}{21.5} & \textbf{4.0} \bigstrut\\
	
	Nikon D3100 & 10 & 720 & 25 & 50 & 100 & $1.6$ & \textcolor{black}{No match} & \textcolor{black}{21.5} & - \bigstrut\\
	
	Nikon D3100 & 11 & 720 & 25 & 75 & 100 & $10.9$ & $\textcolor{black}{22.5}$ & \textcolor{black}{21.5} & \textbf{1.0} \bigstrut\\
	
	Nikon D3100 & 12 & 720 & 25 & 75 & 50 & $3.2$ & $\textcolor{black}{32.5}$ & \textcolor{black}{21.5} & 11.0 \bigstrut\\
		
    \hline\hline
  \end{tabular}
\label{Table: Harmonic vs idle for tested videos}
\end{table*}

\begin{table*}[t]
%\begin{table}[ht]
\caption{\textcolor{black}{Idle Period Estimation Statistics for different cameras for 720 P and 29.97 fps videos}\textcolor{black}{.}}
\centering
  \begin{tabular}{c c c c}
  	\hline\hline
	\textcolor{black}{Camera Model} & \textcolor{black}{Expected idle ($\%$) \cite{Hajj-Ahmad2016-ENF}} & \textcolor{black}{Estimated idle ($\%$)} & \textcolor{black}{Average Estimation Error} \bigstrut\\
	\hline
	\textcolor{black}{GoPro Hero 4} & \textcolor{black}{72.69} & \textcolor{black}{72.50} & \textcolor{black}{1.66} \bigstrut\\
	\textcolor{black}{Nikon D3100} & \textcolor{black}{4.30} & \textcolor{black}{5.00} & \textcolor{black}{4.16} \bigstrut\\
	\textcolor{black}{Nikon P100} & \textcolor{black}{40.55} & \textcolor{black}{40.00} & \textcolor{black}{2.91} \bigstrut\\
	\textcolor{black}{Canon SX230HS} & \textcolor{black}{38.00} & \textcolor{black}{37.50} & \textcolor{black}{2.50} \bigstrut\\
	\textcolor{black}{Canon SX220HS} & \textcolor{black}{40.59} & \textcolor{black}{37.50} & \textcolor{black}{5.00} \bigstrut\\
    \hline\hline
%    \multicolumn{3}{c}{TD = True Decision, FD = False Decision, ND = No Decision} \\
  \end{tabular}
\label{Table:Statistics_for_idle}
\end{table*}

\section{A Novel Time-of-recording Verification Approach}
In this section, a novel time-of-recording verification technique for videos exposed by rolling shutter mechanism is proposed. It is based on a systematic search of possible \textcolor{black}{ENF components} that emerge as a result of idle period, followed by idle period assumptions in each \textcolor{black}{component} and interpolation of missing samples for each assumption.

According to the analytical model introduced in section \ref{AnalyticalModel}, the strongest \textcolor{black}{ENF components} in an ENF containing video can be located. However, for some videos, these \textcolor{black}{ENF components} may contain traces of a non-ENF signal in addition to the ENF. This non-ENF signal may reduce the quality of the ENF signal estimated from these \textcolor{black}{components}. On the other hand, \textcolor{black}{the third strongest ENF component, the fourth strongest ENF component} and so on, whose power is very low compared to the strongest two, may contain a pure ENF trace. In addition to consideration of the other \textcolor{black}{ENF components}, the proposed method also relies on  idle period assumptions and interpolation of missing samples for  each chosen assumption. {Consequently, the use of multiple \textcolor{black}{ENF components}, the idle period assumptions in each \textcolor{black}{component}, followed by interpolation of missing ENF samples for the each assumption can lead to better quality ENF signal estimations. A higher quality of estimated ENF signal generally leads to obtain a higher correlation coefficient value with the ground-truth ENF, hence resulting in a better performing time-of-recording verification. To the best of our knowledge, no state-of-the-art technique including \cite{MinWu-SeeingENFPower-signature-basedtimestampfordigitalmultimediaviaopticalsensingandsignalprocessing}, \cite{su2014}, \cite{patent:20150356992} rely on idle period assumption and interpolation of missing samples for ENF estimation from videos sampled by the rolling shutter. Hence it may be a challenge for these techniques to obtain a good quality of ENF signal for some cases, including compressed videos and videos with moving content.}

%The main operational steps of the proposed time-of-recording verification method are described in Table \ref{Table:Time-of-recording Verification}.
\textcolor{black}{The operational procedure of the proposed time-of-recording verification method consists  of several steps}. First, the time-series for illumination variation throughout the video period is formed via concatenation of the illumination samples of all rows of all frames as in section \ref{Section: Proposed Source Camera Verification}. Following this, the possible \textcolor{black}{ENF components} that may emerge as a result of the idle period are established based on the analytical model. For \textcolor{black}{the each ENF component}, the idle period of the video is presumed to be consecutively 5\%, 10\%, ... 95\%. For the each idle period presumption, missing illumination samples in each frame are interpolated. In reality, it is a great challenge to estimate so many missing samples properly. However, as the ENF signal does not show abrupt changes, an approximation can be made. The interpolation technique used in this work is based on simply averaging {the illumination samples} of the previous and of the next frame. From the each of the interpolated time-series, ENF signal can be estimated by utilization of any time domain or frequency domain techniques discussed in \cite{MinWu-SeeingENFPower-signature-basedtimestampfordigitalmultimediaviaopticalsensingandsignalprocessing}. In this work, ENF is computed via Short Time Fourier Transform (STFT) using 20 seconds time windows with 19 seconds overlaps, followed by quadratic interpolation, which results in 1-second temporal ENF resolution \cite{Vatansever2017_SPL}. Each estimated ENF signal from the each interpolated time-series is compared with the ground-truth ENF data via normalized cross correlation operation. The resulting peak correlation coefficient and the corresponding lag point for each operation are recorded. All the resulting peak correlation coefficients and the corresponding lag points are analysed with the use of an appropriate metric proposed in section \ref{metrics}, i.e.\textcolor{black}{,} Metric 3 or Metric 4. If the final lag point yielded by the applied metric corresponds to the difference between the given time-of-recording and the beginning time of the ground-truth ENF, it is concluded that the given video time-of-recording is true.

\subsection{Experiments with Videos with Still Content}
\label{metrics}
The proposed time-stamp verification algorithm was applied on a wall-scene video dataset that was created by Nikon D3100 and Canon PowerShot SX230HS model cameras in Turkey, where nominal ENF is 50 Hz. With the use of each camera, 2 videos with 29.97 fps in 720 P were taken under each of 3 different light sources, namely LED, Halogen and CFL (Compact Fluorescent) bulbs. Each of 12 original videos were compressed via FFMPEG with different compression ratios (150k, 100k, 50k) as well as YouTube. Accordingly, a total of 48 compressed wall-scene videos were created. The performance evaluation of the proposed method was conducted by using the following metrics: 

\textbf{Metric 1 \cite{MinWu-SeeingENFPower-signature-basedtimestampfordigitalmultimediaviaopticalsensingandsignalprocessing}:} Peak correlation coefficient $(\rho_{H_i}>TH_c)$ at nominal illumination frequency, $H=100 $ Hz, without idle period assumption, $i=\%0$.

\textbf{Metric 2 \cite{su2014}:} Maximum of peak correlation coefficients $(\rho_{H_i}>TH_c)$ obtained for a variety of \textcolor{black}{ENF components}, $H \in \{10, 40, 70, 100, 130, 160, 190, 200\}$ Hz, without idle assumption, $i=\%0$.

\textbf{Metric 3 (Proposed):} Maximum of peak correlation coefficients $(\rho_{H_i}>TH_c)$ obtained for a variety of \textcolor{black}{ENF components}, $H \in \{10, 40, 70, 100, 130, 160, 190, 200\}$ Hz for 30 fps videos recorded in EU, with a variety of idle period assumptions, $i \in \% \{0, 5, 10,…, 95 \}$, and interpolation accordingly.

\textbf{Metric 4 (Proposed):} Maximum of normalized Euclidean distance $(d_g)$ to $(n_{l_g},\rho_{n_{l_g}})$.

\begin{align}
d_g=\sqrt{{n_{l_g}}^2 + {\rho_{n_{l_g}}}^2}
\label{Eq:metric4}
\end{align}

$n_{l_g}$ : Number of lags in a lag group. A lag group is composed of lags being the same in a specified tolerance.

$\rho_{n_{l_g}}$ : The greatest peak correlation coefficient obtained for each $n_{l_g} \quad (\rho_{n_{l_g}}>TH_c)$.

Metric 3 and Metric 4 are described based on the proposed approach in this work. Whereas Metric 1 and Metric 2 are based on the argument in \cite{MinWu-SeeingENFPower-signature-basedtimestampfordigitalmultimediaviaopticalsensingandsignalprocessing} and in \cite{su2014}, respectively, which are provided for comparisons. It should also be noted that $TH_c$ is selected as 0.94 based on empirical analysis.

Table \ref{Table:Experiment: TimeOfRecording with still content - original} provides  estimated results for each metric for the original wall-scene videos. In the table, TD represents true decision rate, FD depicts false decision rate and ND is no decision. As can be seen from the table, each metric shows  100\% true decision rate. However, the performance noticeably changes for the compressed forms of the videos as can be seen in Table \ref{Table:Experiment: TimeOfRecording with still content - compressed}. The performance of Metric 1 and Metric 2 are lower than Metric 3 and Metric 4. Metric 4 outperforms with a true decision rate 70.83\% in relative to 45.83\%, 56.25\% an 68.75\%, respectively for Metric 1, Metric 2 , Metric 3.

\begin{table}[t]
%\begin{table}[ht]
\caption{Time-of-recording verification performance of the presented metrics for original wall-scene videos\textcolor{black}{.}}
\centering
  \begin{tabular}{c c c c c}
  	\hline\hline
	Method & Metric 1 \cite{MinWu-SeeingENFPower-signature-basedtimestampfordigitalmultimediaviaopticalsensingandsignalprocessing} & Metric 2 \cite{su2014} & Metric 3 & Metric 4 \bigstrut\\
	\hline
	TD (\%) & 100 & 100 & 100 & 100 \bigstrut\\
	FD (\%) & 0 & 0 & 0 & 0 \bigstrut\\
	ND (\%) & 0 & 0 & 0 & 0 \bigstrut\\
	
    \hline\hline
    \multicolumn{5}{c}{TD = True Decision, FD = False Decision, ND = No Decision} \\
  \end{tabular}
\label{Table:Experiment: TimeOfRecording with still content - original}
\end{table}

\begin{table}[t]
%\begin{table}[ht]
\caption{Time-of-recording verification performance of the presented metrics for the compressed wall-scene videos\textcolor{black}{.}}
\centering
  \begin{tabular}{c c c c c}
  	\hline\hline
	Method & Metric 1 \cite{MinWu-SeeingENFPower-signature-basedtimestampfordigitalmultimediaviaopticalsensingandsignalprocessing} & Metric 2 \cite{su2014} & Metric 3 & Metric 4 \bigstrut\\
	\hline
	TD (\%) & 45.83 & 56.25 & 68.75 & 70.83 \bigstrut\\
	FD (\%) & 0 & 4.17 & 4.17 & 2.08 \bigstrut\\
	ND (\%) & 54.17 & 39.58 & 27.08 & 27.08 \bigstrut\\
	
    \hline\hline
    \multicolumn{5}{c}{TD = True Decision, FD = False Decision, ND = No Decision} \\
  \end{tabular}
\label{Table:Experiment: TimeOfRecording with still content - compressed}
\end{table}

\subsection{Experiments with Videos with Moving Content}
In this subsection, the performance of the proposed time-of-recording verification algorithm is evaluated with the use of the same dataset described in section \ref{section:Experiment-source camera verification with moving content}, i.e.\textcolor{black}{,} 24 videos with moving content captured in different environments. The same metrics introduced in \ref{metrics} are used to test effectiveness of the proposed approach on these dataset. The only difference in the way the proposed approach is applied for the 25 fps videos is that $\{25, 50, 75, 100, 125, 150, 175, 200 \}$ Hz \textcolor{black}{ENF components} are used instead of $\{10, 40, 70, 100, 130, 160, 190, 200\}$ based on the analytical model proposed in section \ref{AnalyticalModel}. Other than that, all the steps are applied identically. Table \ref{Table:Experiment: TimeOfRecording with moving content} provides the estimated results for each metric. Accordingly, Metric 3 and Metric 4 outperform Metric 1 and Metric 2 for this video dataset as well with true decision rates 79.16\% and 83.33\%, respectively. The true decision rates for Metric 1 and Metric 2 respectively are 62.5\% and 75\%.

Experiments on  videos with both still content and moving content show that exploitation of multiple \textcolor{black}{ENF components}, along with idle period assumptions in each \textcolor{black}{component} and interpolation of missing ENF samples for the each assumption, yield better results than the state-of-the-art. Although the proposed approach leads to comparable outcomes with metric 3 and with metric 4, metric 4 outperforms.

\begin{table}[t]
%\begin{table}[ht]
\caption{Time-of-recording verification performance of the presented metrics for videos with moving content\textcolor{black}{.}}
\centering
  \begin{tabular}{c c c c c}
  	\hline\hline
	Method & Metric 1 \cite{MinWu-SeeingENFPower-signature-basedtimestampfordigitalmultimediaviaopticalsensingandsignalprocessing} & Metric 2 \cite{su2014} & Metric 3 & Metric 4 \bigstrut\\
	\hline
	TD (\%) & 62.50 & 75.00 & 79.16 & 83.33 \bigstrut\\
	FD (\%) & 8.33 & 8.33 & 12.50 & 8.33 \bigstrut\\
	ND (\%) & 29.16 & 16.66 & 8.33 & 8.33 \bigstrut\\
	
    \hline\hline
    \multicolumn{5}{c}{TD = True Decision, FD = False Decision, ND = No Decision} \\
  \end{tabular}
\label{Table:Experiment: TimeOfRecording with moving content}
\end{table}

\section{Conclusion}
In this work, a comprehensive analysis of the rolling shutter mechanism is provided for ENF based video forensics. First, an analytical model is presented illustrating how \textcolor{black}{the frequency of the main ENF harmonic} is replaced with new \textcolor{black}{ENF components} depending on \textcolor{black}{the length of the idle period per frame}. The model {also} reveals that the power of \textcolor{black}{the captured} ENF signal is inversely proportional to idle period length. Hence, \textcolor{black}{in the presence of noise}, videos with long idle periods may be tough for ENF based forensic analysis. Second, with the use of this model, a novel \textcolor{black}{idle period estimation approach is proposed for camera forensics.} {Unlike the state-of-the-art, i.e.\textcolor{black}{,} vertical phase analysis, which relies on alias ENF}, the proposed method is able to operate {also} on videos, where the frame rate is a divisor of nominal ENF.
%\textcolor{blue}{In vertical phase analysis, since one row of the whole frames is treated separately, the sampling rate is as many as \textit{video frame rate}, i.e. lower than Nyquist Criteria. Whereas in the proposed method, each row is treated independently, consequently leading to have a sampling rate as many as \textit{video frame rate $\times$ number of rows.}}
Thirdly, a novel time-of-recording verification technique is introduced, based on utilization of multiple \textcolor{black}{ENF components} as well as idle period assumptions in each \textcolor{black}{component} and interpolation of missing samples. {This approach can result in higher quality ENF signal estimations, consequently leading the time-of-recording verification performance to increase.} Experiments show that the proposed approach outperforms those in the literature.

%as one single row of the entire frames of the video is processed as a separate group,}

\textcolor{black}{Although the proposed idle period estimation method can be used as an auxiliary feature in camera forensics, it cannot be ignored that there may be videos with the same idle period, even if they were captured by different camera models. Besides, the proposed technique} requires some more research to have better  precision. The \textcolor{black}{ENF components} that emerge due to the second and the third illumination harmonics, i.e.\textcolor{black}{,} 200/240 Hz (EU/US) and 300/360 Hz (EU/US) can be taken into consideration. Including these harmonics may also lead to the performance of the proposed time-stamp verification method to improve. Plus, the use of a better interpolation technique, may further augment the performance.

\ifCLASSOPTIONcaptionsoff
  \newpage
\fi



%\bibliographystyle{IEEEtran}

%
%\begin{thebibliography}{1}
%\end{thebibliography}

\bibliographystyle{IEEEtran}
\bibliography{ENF}

 

%\begin{IEEEbiographynophoto}{Ahmet Emir Dirik}


%\begin{NasirMemon}{Nasir Memon}


%\newpage
\begin{IEEEbiography}[{\includegraphics[width=1in,height=1.25in,clip,keepaspectratio]{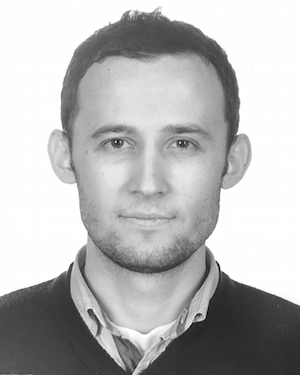}}]{Saffet Vatansever}
received B.Sc. degree in electronics and communications engineering from Yildiz Technical University, Bursa, Turkey, and the M.Sc. degree in mechatronics engineering from University of Newcastle Upon Tyne, Newcastle, UK. He is currently a research assistant with the department of mechatronics, Bursa Technical University, Bursa, Turkey, and  is doing Ph.D. in electronics engineering in Uludag University, Bursa, Turkey. His main research area is multimedia forensics.
\end{IEEEbiography}

\vskip 0pt plus 5fil
%\vskip -0.5\baselineskip plus -1fil

\begin{IEEEbiography}[{\includegraphics[width=1in,height=1.25in,clip,keepaspectratio]{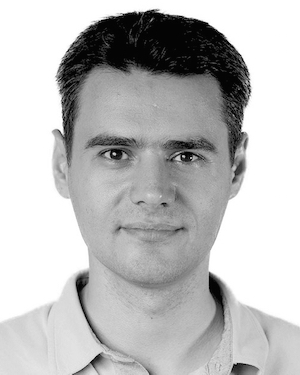}}]{Ahmet Emir Dirik}
received the B.Sc. and M.Sc. degrees in electronics engineering from Uludag University, Bursa, Turkey, and the Ph.D. degree in electrical engineering from the Polytechnic Institute of New York University, Brooklyn, NY, USA, in 2010. He is currently an Associate Professor with the Department of Computer Engineering, Uludag University. His research interests include multimedia forensics and signal processing.
\end{IEEEbiography}

\vskip 0pt plus 5fil
%\vskip -0.5\baselineskip plus -1fil

\begin{IEEEbiography}[{\includegraphics[width=1in,height=1.25in,clip,keepaspectratio]{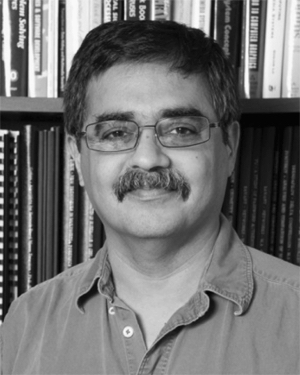}}]{Nasir Memon}
received the M.Sc. and Ph.D. degrees in computer science from the University of Nebraska. He is currently a Professor with the Department of Computer Science and Engineering, New York University (NYU) Tandon School of Engineering, the Director of the OSIRIS Laboratory, a Founding Member of the Center for Cyber Security, a Collaborative Multidisciplinary Initiative of several schools within NYU. He has authored over 250 articles in journals and conference proceedings and holds a dozen patents in image compression and security. His research interests include digital forensics, biometrics, data compression, network security, and usable security. He received several awards, including the Jacobs Excellence in Education Award and several best paper awards. He has been on the editorial boards of several journals and was the Editor-In-Chief of IEEE TIFS. He is a fellow of the IEEE and SPIE. 

 \end{IEEEbiography}


%\vfill


\end{document}